\documentclass[aps,preprint,epsfig,rotate]{revtex4}
\usepackage{graphicx}
\usepackage{bm}
\usepackage{epsfig}

\begin{document}
\title{Photodetachment of the outer-most electron(s) in few-electron atomic systems and variational principles for the  
       cross sections.}

\author{Alexei M. Frolov}
 \email[E--mail address: ]{alex1975frol@gmail.com} .


\affiliation{Department of Applied Mathematics \\
 University of Western Ontario, London, Ontario N6H 5B7, Canada}

\date{\today}

\begin{abstract}

Photodetachment of the outer-most electron(s) in a few-electron atomic systems is investigated. In particular, photodetachment of the 
outer-most electron is considered in the neutral atoms and positively charged atomic ions which contain more than one electron. We also 
consider photodetachment of the outer-most electron in the negatively charged atomic ions, including the negatively charged hydrogen 
H$^{-}$ ion. In all these cases we have derived the closed analytical formulas for the differential and total cross section(s) of 
photodetachment of the outer-most electron(s). For one-electron atoms and ions our method allows one to derive the formulas for the 
photoionization cross sections which exactly coincide with the expressions obtained in earlier studies. A rigorous variational approach, 
which can be applied to determine atomic photoionization and photodetachment cross sections, is also developed. This our approach is 
based on the principle of optimal projection which is the fundamental principle used the physics of transition processes and reactions. \\

\noindent 
PACS number(s): 32.80.Fb, 32.80.Gc and 32.90.+a

\end{abstract}

\maketitle


\section{Introduction}

Currently, there is a deep misconception that analytical methods for dealing with atomic photoionization can be applied only to one-electron atoms 
and ions. In reality, this statement is not true and one eaily finds many photoionization and photodetachment problems which can be solved to very 
good accuracy, or even exactly, by using relatively simple analytical methods and better understanding of the physics of this problem. One of such 
problems is photodetachment of the outer-most electron(s) in few-electron atomic systems which has a great interst in numerous applications, 
including stellar astrophysics, physics of hot plasmas, etc. In this communication by using analytical methods we investigate photodetachment of 
the outer-most electron(s) in atomic systems which include few- and many-electron atoms and ions. The main goal of this study is derivation of the 
explicit formulas for the both differential $\frac{d \sigma}{d o}$ and total $\sigma$ cross sections of photoionization and/or photodetachment as 
the explicit functions of $I, \omega$ and a few other parameters. Here $I$ is the atomic ionization potential, while $\omega$ is the cyclyc 
frequency of the incident light. It should be emphasized that in this study we restrict ourselves to the photodetachment of the outer-most electron 
in few-electron atomic systems. It is assumed $a$ $priori$ that such an electron is weakly bound and its photodetachment has a number of specific 
features, the study of which is the main purpose of our paper. The general theory of atomic photoionization (see, e.g., \cite{Star}, \cite{Sobelman}
and \cite{BS}) and recent computational and experimental results (see, e.g., \cite{Zat}, \cite{Peg} and \cite{Kry}) are mentioned below only to 
explain some details and results of our analysis. The same statement is true for various theories which explain some tiny differences between 
numerical results obtained for photodetachment of the negatively charged atomic ions. However, a number of papers which are important for our 
analysis of photodetachment of the negatively charged ions are used below \cite{Ajm}, \cite{FroF} (see, also \cite{Frit}).     

In this communication we investigate photodetachment of the outer-most electron(s) in atomic systems which include few- and many-electron 
atoms and ions. Our main goal in this study is to obtain the explicit analytical formulas for the both differential and total cross-sections of 
photodetachment of the outer-most electrons in light few-electron atoms and ions. As follows from numerous experiments and asrophysical 
observations such a photodetachment essentially depends upon the atomic ionization potential $I$, which is an unknown function of the total 
number of bounded electrons $N_e$ in the initial atom/ion and electrical charge $Q$ of the atomic nucleus. The second parameter of the problem 
is the frequency, or cyclic frequency $\omega$, of the incident light quantum. In reality, there are three different experimental situations 
which are equally important in applications (see, e.g., \cite{Sobelman}). The first and simplest group of photoionization and photodetachment 
problems includes all one-electron atoms and ions for which $N_e = 1$ ($Q \ge 1$ is an arbitrary integer number). The explicit formulas for the 
both differenetial and total cross sections of photoionization of the ground state in hydrogenic atoms and ions were derived in 1930's \cite{Stob}. 
Later attempts to obtain analogous formulas for the excited states in hydrogenic systems were less successfull and many results for such states 
are wrong by the factor of 2 (and even 4) (see discussion in Sections 69 - 75 of \cite{BS}). Nevertheless, since 1930's the Stobbe formula was 
extensively used to explain and describe workability of various physical systems and devices where the atomic photoionization and photodetachment 
play a crucial role. Such systems/devices included the both atomic and hydrogen bombs (some examples are described, e.g., in \cite{Rodes}), 
stellarators, tokamaks, etc. 

Unfortunately, for few-electron atomic systems, including the neutral atoms and positively charged ions, nobody could ever produce any closed 
analytical formula for the photoionization cross sections. The same situation can also be found for the negatively charged atomic ions, e.g., 
for the H$^{-}$ and Li$^{-}$ ions which are well known and stable atomic ions. Currently, we have a number of numerical methods, which were 
develop to approximate photoionization cross sections for all these systems, and 99.9 \% of these methods are based on the Stobbe formula(s) for 
one-electron atomic systems. In \cite{Bates} some universal procedure to approximate the photoionization cross sections known for various few- 
and many-electron atomic systems. That method was also created with the use of Stobbe's formula(s) as its only source. Later that procedure was 
modified in \cite{BurSeat}. Since then this procedure is extensively used as some standard approach to evaluate the photoionization cross sections 
for many atoms and ions, including negatively charged atomic ions. However, it is $a$ $priori$ clear that photoionization of few- and many-electron 
atoms and positively charged ions cannot be described (completely and accurately) by the Stobbe formula. For photodetachment of the negatively 
charged ions the situation is even worse and applications of Stobbe's formula for such systems always lead to qualitative mistakes. 

The main goal of this study is to derive the closed analytical formulas for the cross sections of photodetachment of the outer-most electrons in 
few-electron atoms and ions. First, we solve this problem for neutral atoms and in positively charged ions which contain few electrons, i.e., in 
atomic systems where $Q \ge N_e \ge 2$. The closed analytical formulas for the cross sections derived below can be applied to describe 
photodetachment of the outer-most electrons in such atomic systems. The second fundamental goal is to produce analytical formulas for the cross 
sections of photodetachment of the outer-most electrons in negatively charged  atomic ions, where $Q = N_e + 1 \ge 2$. This problem has never 
been solved (even approximately) in earlier studies. To achieve these goals we have applied a very accurate (analytical) representation of the 
electron density at large and very large distances from the central positively charged atomic nucleus which is provided by the modern atomic DFT 
theory \cite{Osten}. This means that we have the correct long-range assymptotics of the truly correlated wave functions known for all few- and 
many-electron atomic systems. On the other hand, photodetachment of the outer-most electrons almost always proceeds at large and very large 
distances from the central atomic nucleus. Therefore, we can expect that our method will work better than procedures based on unjustifiable 
numerical approximations of the truly correlated atomic wave functions by one-electron, model functions. Briefly, in this study we want to derive 
analytical formulas for the photodetachment and/or photoionization cross sections of the few-electron atoms and ions, and, in some sense, these
our formulas can be considered as direct analogue of the Strobbe's formula found earlier for one-electron atomic systems. 

Technically, in this study we want to derive the explicit formulas for the both differential $\frac{d \sigma}{d o}$ and total $\sigma$ cross 
sections of photoionization and/or photodetachment as the explicit functions of $I$ and $\omega$. In reality, for photoionization cross sections of 
the neutral few-electron atoms and positively charged ions we obtain the first group of formulas which represent the $\sigma(I, \omega)$ dependencies. 
A different form of the the $\sigma(I, \omega)$ functions is derived for the photodetachment cross sections of the negatively charged ions. The third 
group of analytical fromulas for the $\sigma(I, \omega)$ cross sections describes photoionization of one-electron atomic systems. One-electron atomic 
systems are considered in our analysis only for methodological purposes to check our method. Furthermore, these formulas are also useful to check our 
newly derived formulas in the limit $N_e = 1$ (they must coincide with the Stobbe formula(s)). In all these cases, our formulas derived for the both
differential and total cross sections of photodetachment are written in the closed and explicit forms. Some applications of these formulas to actual 
astrophysical systems are also discussed. We have also developed the new variational approach which can be applied to increase the overall accuracy 
and flexibility of many modern numerical methods which are used currently to determine photoionization and photodetachment cross sections of various 
atomic systems. 

\section{Astrophysical applications}
 
Different solutions of the photoionization and/or photodetachment problem are of great interest in a large number of applications which include 
photoionization of actual atoms and ions arising in various stellar and laboratory plasmas. In stellar astrophysics photodetachment of the 
outer-most electrons in few-electron neutral atoms and positively charged ions is very important to understand and describe opacities of different 
stellar photoshperes. In general, such opacities depend upon the temperature and pressure at the star's surface, pre-surface gradietns of these 
values and some other factors. For instance, for very hot $O-$stars, where the surface temperatures $T_s$ exceed 75,000 $K$ - 90,000 $K$ (and even 
125,000 $K$ for some stars), we have to deal with photoionization of the neutral, two-electron He atoms and positively charged, one-electron 
He$^{+}$ ions. Opacities of the stellar photospheres of less hotter $B$ stars, where the surface temperatures $T_s$ vary between $\approx$ 18,000 
$K$ and $\approx$ 27,000 $K$, are mainly determined by photoionization of the neutral hydrogen atoms and optical transitions in them.  

A separate group of the hot $Be-$stats is of great interest to study photoionization of hydrogen atoms from its different ground and excited 
states. The $Be-$stars are typical $B-$stars which also have rapidly rotating disks (or rings) of hydrogen atoms and molecules (see, e.g., 
\cite{Marl} and references therein). The well known $Be$-star is $\gamma-$Cassiopeiae (or $\gamma-$Cas, for short) which is relatively close 
to our Sun. This star is easy to find at the clear night sky in the northern hemisphere, since it is located at the central vertex of the 
distinctive W-shape in the northern circumpolar constellation of Cassiopeiae. In all Be-stars, including $\gamma-$Cas and other similar 
stars, the hot central star illuminates surrounding rings of low-dense hydrogen atoms which rotate at different distances around the central, 
very hot $B-$star. Stellar radiation is absorbed by these hydrogen atoms, which rapidly move in the ring (their typical velocities are $v_{H} 
\approx$ 300 - 500 $km \cdot sec^{-1}$). Then the absorbed radiation is re-emitted (with smaller frequencies) by these rapidly moving hydrogen 
atoms in all directions. An observer from a distance of a few hundreds of light years can see the regular (or thermal) spectrum of the central 
$B-$star, which is combined with one (or a few) sharp and very intense spectral lines which describes the optical transitions in neutral hydrogen
atoms. This explains presence of very intense red hydrogen line(s) in the emmision spectrum of $\gamma-$Cassiopeiae. Since any of these $Be-$stars 
contains hydrogen atoms at different temperatures, then we shall also observe photodetachment of these neutral atoms from different excited (and 
ground) states, including weakly-bound (or Rydberg) states.  

Photodetachment of the negatively charged hydrogen H$^{-}$ ions plays a great role in understanding of the actual optical and infrared spectra of 
many late $F-$ and all $G-$stars (see discussion and references in \cite{Fro2015A}). Our Sun is a star which belongs to the $G2$ spectral class 
and its average surface temperature is $\approx$ 5,773 $K$. Formally, if photodetachment of the negatively charged hydrogen ions in Solar 
photosphere is ignored, then it is impossible to explain the actual emission spectrum of our Sun located between $\lambda_{min} \approx$ 6500 $\AA$ 
and $\lambda_{max} \approx$ 20,000 $\AA$, i.e., we cannot accurately describe neither optical, nor infrared spectra at these wavelengths \cite{Sob} 
- \cite{Zirin}. 

\section{Photoionization and photodetachment cross sections for atoms and ions}

In general, photoionization and/or photodetachment of any atomic system is defined as absorption of a photon by some bound atomic electron. Such an 
absorbtion produces an instant transition of this initial electron into the final unbound state (or state of unbound spectrum). The arising unbound 
(or 'free') electron moves away from the parental atom/ion in the Coulomb field of the final atomic ion. In the case of negatively charged H$^{-}$ 
ion the unbound photoelectron moves in the field of a neutral hydrogen atom. As follows from arguments presented above in order to solve various 
astrophysical problems one needs to know the closed analytical expressions for the cross sections of photodetachment of the outer-most electrons in 
different atomic systems including one-, few- and many-electron atoms and positively charged ions. On the other hand, it is important to produce the 
closed analytical formulas for the photodetachment cross sections of the outer-most electrons in a number of negatively charged atomic ions. In 
earlier studies the closed expression for the photoionization cross section was produced by Stobbe \cite{Stob} only for the ground states in 
one-electron atoms and ions (see Appendix).  

Note that the general theory of atomic photoionization and photodetachment is a well developed chapter of modern Quantum Electrodynamics (see, e.g., 
\cite{AB}). Formulas for the differential and total cross sections of photodetachment were derived and discussed in numerous QED textbooks (see, 
e.g., \cite{AB} - \cite{Grein}). This fact allows us to reduce the introductory (or QED) part of this study to a minimum and proceed directly to 
derivation of important formulas and discussion of our original results. According to the rules of modern Quantum Electrodynamics (see, e.g., 
\cite{AB} - \cite{Grein}) the differential cross section of photoionization and/or photodetachment of an arbitrary atomic system is written in the 
following form \cite{AB}  
\begin{eqnarray}
 d\sigma = \frac{e^{2} m \mid {\bf p} \mid}{2 \pi \omega} \mid {\bf e} \cdot {\bf v}_{fi} \mid^{2} do = \frac{p}{2 \pi \omega a_0} 
 \mid {\bf e} \cdot {\bf v}_{fi} \mid^{2} do \; \; \; \label{cross}
\end{eqnarray}
where $m \approx$ 9.1093837015$\cdot 10^{-28}$ $g$ is the rest mass of electron, $a_0 \approx$ 5.29177210903 $cm$ is the Bohr radius, while 
$- e$ is its electric charge and ${\bf e}$ is the vector which describes the actual polarization of initial photon. Also, in this formula 
$p = \mid {\bf p} \mid$ is the momentum of the final (or free) photoelectron, $\omega$ is the cyclic frequency of the incident light quanta, 
while ${\bf v}_{fi}$ is the matrix element of the `transition' velocity ${\bf v}_{fi}$. For this matrix element we can write ${\bf v}_{fi} = 
- \frac{\imath}{m} \langle \psi_{f} | \nabla | \psi_{i} \rangle$, where the notation $\psi$ designates the wave functions, while the indexes 
$f$ and $i$ in this equation and in all formulas below stand for the final and initial states, respectively. The formula, Eq.(\ref{cross}), 
is written in the relativistic units, where $\hbar = 1, c = 1, e^{2} = \alpha$ and $\alpha \approx$ 7.2973525693$\cdot 10^{-3}$ ($\approx 
\frac{1}{137}$) is the dimensionless fine-structure constant. These relativistic units are convenient to perform analytical calculations in 
Quantum Electrodynamics (below QED, for short). However, in order to determine the non-relativistic cross-sections it is better to apply 
either the usual units, e.g., $C G S$ units, or atomic units in which $\hbar = 1, m_e = 1$ and $e = 1$. In these atomic units the formula, 
Eq.(\ref{cross}), takes the form
\begin{eqnarray}
 d\sigma = \frac{\alpha a^{2}_{0} p}{2 \pi \omega} \mid {\bf e} \langle \psi_{f} | \nabla | \psi_{i} \rangle \mid^{2} do \; \; \; 
 \label{crossau}
\end{eqnarray}
where $a_{0} = \frac{\hbar^{2}}{m e^{2}}$ is the Bohr radius and $\alpha = \frac{e^{2}}{\hbar c}$ is the dimensionless fine-structure constant. 
In atomic units we have $a_{0} = 1$ and $\alpha = \frac{1}{c} (\approx \frac{1}{137})$. Let us assume that initial electron was bound to some 
atomic system, i.e., to a neutral atom, negatively and/or positively charged ion. The energy of this bound state (or discrete level) is 
$\varepsilon = - I$, where $I$ is the atomic ionization potential. It is clear that the condition $\omega \ge I$ which must be obeyed to make 
photoionization and/or photodetachment possible. Here and everywhere below in this study the notation $I$ stands for the ionization potential 
of the initial atomic system and we have the following relation $\omega = I + \frac12 p^{2}$ which is also written in the form $p = \sqrt{2 
(\omega - I)}$. In this study the transition matrix element in Eqs.(\ref{cross}) and (\ref{crossau}) is written in the velocity form (see, e.g., 
\cite{BD1}). This form directly follows from Quantum Electrodynamics. Other possible forms of the transition matrix element, e.g., its length 
form \cite{BD1}, are not used below.  
  
\subsection{Wave functions of the final and initial states}
 
Now, we need to develop a logically closed procedure to calculate the matrix element (or transition amplitude) $\langle \psi_{f} | \nabla 
| \psi_{i} \rangle$ which is included in the formulas, Eqs.(\ref{cross}) and (\ref{crossau}). Everywhere below in this study, we shall 
assume that the original (or incident) system was in its lowest-energy ground (bound) state. For one-electron atomic system this state is 
always the doublet $1^{2}s(\ell = 0)-$state. Then, in the lowest order dipole approximation \cite{AB} the outgoing (or final) photoelectron 
will move in the $p(\ell = 1)-$wave. By using this fact we can write the following expression for the wave function of the final electron
\begin{eqnarray}
 \psi_{f}({\bf r}) = \frac{2 \ell + 1}{2 p} \; \; P_{\ell}({\bf n} {\bf n}_1) \; \psi_{\ell;p}(r) = ({\bf n} {\bf n}_1) \; \;
 \frac{3}{2 p} \psi_{1;p}(r) = ({\bf n} {\bf n}_1) R_{1;p}(r) \; \; \label{psi-f}
\end{eqnarray} 
where $\ell = 1, P_{\ell}(x)$ is the Legendre polynomial (see, e.g., \cite{GR}) and $R_{\ell=1;p}(r) = R_{1;p}(r)$ is the corresponding 
radial function which depends upon the explicit form of the interaction potential between outgoing photoelectron and remaining atomic 
system. Also, in this formula the unit vectors ${\bf n}$ and ${\bf n}_1$ are: ${\bf n} = \frac{{\bf p}}{p}$ and ${\bf n}_1 = \frac{{\bf 
r}}{r}$. The unit vector ${\bf n}$ determines the direction of outgoing (or final) photoelectron, or ${\bf p}-$direction, for short. 

If the interaction potential between outgoing photoelectron and remaining atomic system is described by a Coulomb potential, then the 
radial function $\psi_{\ell;p}(r)$ in Eq.(\ref{psi-f}) is the normalized Coulomb function of the first kind (see, e.g., \cite{AS}) which 
is 
\begin{eqnarray}
 \psi_{1;p}(r) = \frac{2^{\ell} Z}{(2 \ell + 1)!} \sqrt{\frac{8 \pi}{\nu [1 - \exp(-2 \pi \nu)]}} \; \; \; 
 \prod^{\ell}_{s=1} \sqrt{s^{2} + \nu^{2}} \; \; (p r)^{\ell} \; \exp(-\imath p r) \; \times \nonumber \\
 \; {}_1F_{1}(\ell + 1 + \imath \nu, 2 \ell + 2; 2 \imath p r) \; \; \label{radiala0}
\end{eqnarray} 
where ${}_1F_{1}(a, b; z)$ is the confluent hypergeometric function defined exactly as in \cite{GR} and \cite{AS}, $\nu = \frac{Z}{p} = 
\frac{Q - N_e + 1}{p}$ and $Z = Q - N_e + 1$ is the electric charge of the final atomic fragment, e.g., electrical charge of a bared 
nucleus for one-elecron atomic system. From this equation one finds for $\ell = 1$
\begin{eqnarray}
 \psi_{1;p}(r) &=& \frac{2 Z p}{3!} \sqrt{\frac{8 \pi (1 + \nu^{2})}{\nu [1 - \exp(-2 \pi \nu)]}} \; \; r \exp(-\imath p r) \; \; 
 {}_1F_{1}(2 + \imath \nu, 4; 2 \imath p r) \; \; \label{radiala}
\end{eqnarray} 
By multiplying this formula by the additional factor $\frac{3}{2 p}$ from Eq.(\ref{psi-f}), we can write for $\ell = 1$ (in atomic units) 
\begin{eqnarray}
 R_{1;p}(r) &=& Z \sqrt{\frac{2 \pi (1 + \nu^{2})}{\nu \Bigl((1 - \exp(-2 \pi \nu)\Bigr)}} \; \; r \exp(-\imath p r) \; \; {}_1F_{1}(2 + 
 \imath \nu; 4; 2 \imath p r) \; \; \nonumber \\
  &=& p \sqrt{\frac{2 \pi \nu (1 + \nu^{2})}{\Bigl((1 - \exp(-2 \pi \nu)\Bigr)}} \; \; r \exp(-\imath p r) \; \; {}_1F_{1}(2 + 
 \imath \nu; 4; 2 \imath p r) \; \; \label{rad1}
\end{eqnarray} 
where the electric charge $Z$ of the remaining atomic system is an increasing function of the nuclear charge $Q$, but it also depends upon 
the total number of bound electrons $N_e$. All phases and normalization factors in this formula are chosen exactly as in \cite{Fock} and 
\cite{LLQ}. 

For the non-Coulomb (or short-range) interaction potentials between outgoing photoelectron and remaining atomic cluster the normalized radial 
wave function of the continous spectra is written as a product of the spherical Bessel function $j_{\ell}(p r)$ and a factor which equals $2 
p$, i.e., $\psi_{\ell;p}(r) = 2 p j_{\ell}(p r)$ (see, e.g., \cite{LLQ}, \$ 33). For $\ell = 1$ one finds $\psi_{1;p}(r) = 2 p j_{1}(p r)$. 
This means that the function $\psi_{\ell=1;p}(r)$ is regular at $r = 0$. From here for the radial function $R_{1;p}(r) = \frac{3}{2 p} 
\psi_{\ell=1;p}(r)$ we obtain 
\begin{eqnarray}
 R_{1;p}(r) = \frac{3}{2 p} \Bigl[ 2 p j_{\ell=1}(p r) \Bigr] = 3 j_{\ell=1}(p r) = 3 \sqrt{\frac{\pi}{2 p r}} \; \; J_{\frac32}(p r) \; 
 \; \label{rad2}
\end{eqnarray}
where $j_{1}(x) = \frac{sin x}{x^{2}} - \frac{cos x}{x}$ and $J_{\frac32}(x)$ is the Bessel function which is regular at the origin (at $r = 
0$) and defined exactly as in \cite{GR}.
 
As mentioned above our goal in this study is to derive analytical formulas for the photodetachment cross sections in a number of actual cases 
which differ from each other by the choice of the initial and/or final electron's wave function(s). The two possible wave functions of the 
final unbound electron are mentioned above. Now, let us discuss the wave functions of the initial atomic system which has one nucleus with 
the electric charge $Q$ (or $Q e$) and $N_e$ bound electrons. By analyzing the current experimental data it is easy to understand that the 
non-relativistic photodetachment of the outer-most electrons in a few- and many-electron atomic systems is produced by photons with large and 
very large wavelengths $\lambda$. In reality, the wavelengths $\lambda$ of incident light quanta substantially exceed the actual sizes of 
atoms and ions. For instance, the wavelengths $\lambda$ of photons that produce photodetachment of the negatively charged hydrogen ions 
H$^{-}$ in Solar photosphere exceed 7000 $\AA$ (or $\ge $ 13232 $a.u.$), while the spatial radius $R$ of this ion equals $\approx$ 
2.710$\ldots$ $a.u.$ (see e.g., \cite{Fro2015}), i.e., $R \ll \lambda$ (in fact, $R$ is smaller than $\lambda$ in thousands times!). In other 
words, photodetachment of the outer-most electron in the H$^{-}$ ion is produced at large and very large distances from the central atomic 
nucleus and from the rest of atomic elecrons. Such asymptotic spatial areas in the H$^{-}$ ion are very important to determine photodetachment 
cross sections, since only in these spatial areas one finds a relatively large overlap between the electron and photon wave functions. 

Similar situations can be found for other atomic systems considered in this study, e.g., for all neutral atoms and positively charged ions. 
In each of these Coulomb systems photodetachment of the outer-most electron(s) mainly occurs in the asymptotic areas of their wave functions. 
These asymptotic areas are located far (even very far) form the central atomic nucleus and other internal atomic electrons. Therefore, in our 
analysis of non-relativistic photodetachment of the outer-most electron(s) we can restrict ourselves to large spatial areas and consider only 
the long-range asymptotics of these wave functions. Moreover, it seems very tempting to neglect the small area of electron-electron correlations 
around the central nucleus and consider the long-range asymptotics of atomic wave functions as the `new' wave functions for our problem. Briefly, 
in this procedure we replace the actual wave functions for each of these atomic systems by their long-range radial asymptotics. It is clear that 
the `new' bound state wave function is one-electron function and it has a different normalization constant. Obviously, this is an approximation,  
but as follows from our results the overall accuracy of our approximation is very good and sufficient to describe photodetachment of the outer 
most-electrons in all atomic systems discussed in this study. 

For one-electron atoms and/or ions the radial asymptotics of the true wave function coincides with the radial wave function itself. Therefore, for 
such systems our method is not an approximation, but an exact method of solution. For neutral atoms and positively charged ions our method is 
accurate enough to represent all qualitative features of the exact wave function and obtain a good quantitative approximation for it. At least it
is sufficient to solve our photodetachment problem to the accuracy $\approx$ 4 - 7 \%. For the negatively charged atomic ions the actual situation 
is worst, but even in these cases the maximal numerical error in photodetachment cross sections does not exceed 10 \% - 15 \%. This fact was checked
many times for different atomic systems by comparing photodetachment cross sections of different negatively charged ions with the known experimental 
cross sections and results of more accurate numerical computations. In reality, by varying a few additional parameters (usually, one, two and three 
parameters) we can approximate numerical cross section to good and even very good accuracy (for more details, see, e.g., \cite{Bet2}). This simple 
technique is desribed below.  
      
In this study we consider the three following atomic systems: (a) atom/ion which initially contains $N_e$ bound electrons, while its nuclear 
charge $Q$ ($Q \ge N_e$) is arbitrary, (b) one-electron atom/ion, where $N_e = 1$ and $Q$ is arbitrary, and (c) negatively charged ion where 
$Q = N_e + 1$ and the both $Q$ and $N_e$ are arbitrary. As is well known (see, e.g., \cite{Osten} and references therein) in arbitrary atomic 
$(Q, N_e)$-system the radial wave function of the ground $S(L = 0)$-states has the following long-distance (radial) asymptotics 
\begin{eqnarray}
 R_{i}(r) &=& C(b;I) r^{b - 1} \exp(- \sqrt{2 I} r) = \frac{(2 \sqrt{2 I})^{b + \frac12}}{2 \sqrt{\pi \Gamma(2 b + 1)}} \; \; r^{b - 1} 
 \; \exp(- \sqrt{2 I} r) \; \nonumber \\
 &=& \frac{(2 \sqrt{2 I})^{b + \frac12}}{2 \sqrt{\pi \Gamma(2 b + 1)}} \; \; r^{\frac{Q - N_e + 1}{\sqrt{2 I}} - 1} \; 
 \exp(- \sqrt{2 I} r) \; \; \label{rad3}   
\end{eqnarray}
where $b = \frac{Q - N_e + 1}{\sqrt{2 I}} = \frac{Z}{\sqrt{2 I}}, I (\ge 0)$ is the atomic ionization potential and $Z = Q - N_e + 1$ (or 
$Z = (Q - N_e + 1) e$) is the electric charge of the remaining atom/ion. In Eq.(\ref{rad3}) and everywhere below the notation $\Gamma(x)$ 
denotes the Euler's gamma-function $\Gamma(1 + x) = x \Gamma(x)$, which is often called the Euler's integral of the second kind \cite{GR}.
This important result of Density Functional Theory (or DFT, for short) plays a central role in this study. Here we have to emphasize the 
following fundamental fact: the formula, Eq.(\ref{rad3}), is the exact long-range asymtotics of the truly correlated, $N_e$-electron wave 
function of an actual atom/ion, and it is not based on any approximation. In other words, by choosing the wave function $R_{i}(r)$ in the 
form of Eq.(\ref{rad3}) we do not neglect any of the electron-electron correlations in atomic wave function. Since photodetachment of the 
outer-most electrons mainly occurs at large and very large distances from the atomic nucleus, then it will be a very good approximation to
describe this phenomenon, if we continue the radial $R_{i}(r)$ function, Eq.(\ref{rad3}), on the whole real $r-$axis, including the radial 
origin, i.e., the point $r = 0$. This allows us to determine the factor $C(b;I)$ in Eq.(\ref{rad3}) which is the normalization constant of 
the radial $R_{i}(r)$ function, which now continues on the whole real $r-$axis, including the radial origin, i.e., the point $r = 0$. This 
normalization constant for this radial $R_{i}(r)$ function, Eq.(\ref{rad3}) where $0 \le r < +\infty$, equals  
\begin{eqnarray}
   C(b;I) = \frac{(2 \sqrt{2 I})^{b+\frac12}}{2 \sqrt{\pi \Gamma(2 b + 1)}} \; \; \label{norm}
\end{eqnarray}
where the atomic ionization potential $I$ and 'power parameter' $b$ are the two real, non-negative numbers. In some places below the $\sqrt{2 
I}$ value is also designated as $B$. In the general case, the atomic ionization potential $I$ is an unknown function of $Q$ and $N_e$. 

For the negatively charged (atomic) ions we always have $b = \frac{Q - N_e + 1}{\sqrt{2 I}} = 0$. This means that the long-distance asymptotic 
of the radial wave function of an arbitrary negatively charged ion is $R(r) \sim \frac{C}{r} \exp(-\sqrt{2 I} r)$, where $C = 
\sqrt[4]{\frac{I}{2 \pi^2}}$ is the normalization constant and $I$ is an unknown function of $Q$ and $N_e$. In contrast with this, for 
one-electron atoms and ions we have $N_e = 1$ and $2 I = Q^{2}$, if photodetachment of the ground atomic state is considered. In this case the 
ionization potential $I$ is the uniform function of $Q$ only. For one-electron atomic systems we have $b = 1$ and the exact wave function is
written in the form $R(r; Q) = A \exp(- \sqrt{2 I} r)$, where $2 I = Q^{2}$, and $A = \frac{Q \sqrt{Q}}{\sqrt{\pi}}$ is the normalization 
constant. This case is important in numerous applications to actual one-electron atoms and positively charged (or multi-charged) ions.  

\subsection{Gradient operator and its matrix elements}

Let us derive some useful formulas for the matrix element $\langle \psi_{f} | \nabla | \psi_{i} \rangle$ which is inlcuded in Eq.(\ref{crossau}). 
It is clear that we need to determine the vector-derivative (or gradient) of the initial wave function, which is a scalar function. In general, 
for the interparticle (or relative) vector ${\bf r}_{ij} = {\bf r}_{j} - {\bf r}_{i}$ the corresponding gradient operator in spherical coordinates 
takes the form (see, e.g., \cite{Kochin} and \cite{Jacks})
\begin{eqnarray}
 \nabla_{ij} = \frac{d }{d {\bf r}_{ij}} = \frac{{\bf r}_{ij}}{r_{ij}} \frac{\partial }{\partial r_{ij}} + \frac{1}{r_{ij}} \nabla_{ij}(\Omega) 
 = {\bf e}_{r;ij} \frac{\partial }{\partial r_{ij}} + {\bf e}_{\theta;ij} \frac{1}{r_{ij}} \frac{\partial }{\partial \theta_{ij}} 
 + {\bf e}_{\phi;ij} \frac{1}{r_{ij} \sin\theta_{ij}} \frac{\partial }{\partial \phi_{ij}} \; \; \label{grad}
\end{eqnarray}
where $\nabla_{ij}(\Omega)$ is the angular part of the gradient vector which depends upon angular variables ($\theta$ and $\phi$) only, while 
${\bf e}_{r;ij} = \frac{{\bf r}_{ij}}{r_{ij}} = {\bf n}_{ij}, {\bf e}_{\theta;ij}$ and ${\bf e}_{\phi;ij}$ are the three unit vectors in 
spherical coordinates which are defined by the ${\bf r}_{j}$ and ${\bf r}_{i}$ vectors, where ${\bf r}_{j} \ne {\bf r}_{i}$. More useful 
formulas for the gradient operator(s), written in spherical coordinates, can be found, e.g., in \$ 5.7 of the excelent book by Edmonds 
\cite{Edm}. 

If the radial part of the initial wave function depends upon the scalar radial variable only, then all derivatives in respect to the both 
angular variables $\theta$ and $\phi$ equal zero identically and we can write    
\begin{eqnarray}
 \nabla_{ij} R(r) = \frac{{\bf r}_{ij}}{r_{ij}} \frac{\partial R(r_{ij})}{{\partial r}_{ij}} = \frac{{\bf r}_{ij}}{r_{ij}} \frac{d 
 R(r_{ij})}{d r_{ij}} = {\bf n}_{ij} \frac{d R(r_{ij})}{d r_{ij}} \; \; \label{grad1}
\end{eqnarray} 
where ${\bf n}_{ij}$ is the unit vector in the direction of interparticle ${\bf r}_{ij}$ variable. For one-center atomic systems we can 
determine $r_{1j} = r_{j}$, and for one-electron systems $r_{12} = r_1 = r$. In this case, the formula, Eq.(\ref{grad1}), is written in the 
form: $\nabla R(r) = {\bf n}_1 \frac{d R(r)}{d r}$, where ${\bf n}_1 = \frac{{\bf r}_{1}}{r_{1}}$. In this notation the radial matrix 
element is 
\begin{eqnarray}
 & & {\bf e} \langle \psi_{f} | \nabla | \psi_{i} \rangle = \int_{0}^{+\infty} \Bigl\{\oint ({\bf n} \cdot {\bf n}_1) 
 ({\bf e} \cdot {\bf n}_1) do_1 \Bigr\} \Bigr( R_{1;p}(r) \frac{d R_{i}(r)}{d r} \Bigl) r^{2} dr \nonumber \\
 &=& \frac{4 \pi}{3} ({\bf e} \cdot {\bf n}) \int_{0}^{+\infty} \Bigr( R_{1;p}(r)\frac{d R_{i}(r)}{d r} \Bigl) r^{2} dr = 
 \frac{4 \pi}{3} ({\bf e} \cdot {\bf n}) I_{rd} \; \; \label{ME}
\end{eqnarray} 
where $R_{1;p}(r)$ and $R_{i}(r)$ are the radial functions of the final and initial states, respectively. The notation $I_{rd}$ in this 
formula, Eq.(\ref{ME}), stands for the following auxiliary radial integral
\begin{eqnarray} 
 I_{rd} = \int_{0}^{+\infty} \Bigr( R_{1;p}(r) \frac{d R_{i}(r)}{d r} \Bigl) r^{2} dr = - \int_{0}^{+\infty} \Bigr( R_{i}(r) 
 \frac{d R_{1;p}(r)}{d r} \Bigl) r^{2} dr \; \; \label{aux}
\end{eqnarray} 
where we used the so-called 'transfer of the derivative' (or partial integration) which often helps to simplify analytical calculations of 
this radial integral. 

By substituting the expression, Eq.(\ref{ME}), into the formula, Eq.(\ref{crossau}), one finds the following 'final' formula for the 
differential cross section of the non-relativistic photodetachment of an arbitrary atomic system
\begin{eqnarray}
 d\sigma = \frac{16 \pi^{2}\alpha a^{2}_{0} p}{18 \pi \omega} ({\bf e} \cdot {\bf n})^{2} \mid I_{rd} \mid^{2} do 
 = \frac{8 \pi\alpha a^{2}_{0} p}{9 \omega} ({\bf e} \cdot {\bf n})^{2} \mid I_{rd} \mid^{2} do \; \; \label{sigma-a}
\end{eqnarray} 
As follows from this formula the angular distribution of photoelectrons is determined by the 'angular' factor $({\bf e} \cdot {\bf n})^{2}$. 
This cross section of photodetachment corresponds to the truly (or 100 \%) polarized light. However, in many actual applications the 
incident beam of photons is unpolarized and we deal with the natural (or white) light. If the incident beam of photons was unpolarized, then 
we need to apply the formula $\overline{({\bf e} \cdot {\bf n})^{2}} = \frac12 ({\bf n}_l \times {\bf n})^{2}$, where ${\bf n}_l$ is the unit 
vector which describes the direction of incident light propagation and ${\bf n}$ is the unit vector which determines the direction of 
propagation of final photoelectron. In this study the notation $({\bf a} \times {\bf b})$ denotes the vector product of the two vectors ${\bf 
a}$ and ${\bf b}$. Finally, the differential cross section of photodetachment takes the form 
\begin{eqnarray}
 d\sigma = \Bigl(\frac{4 \pi \alpha a^{2}_{0} p}{9 \omega}\Bigr) \; ({\bf n}_{l} \times {\bf n})^{2} \mid I_{rd} \mid^{2} ({\bf n}_l \times 
 {\bf n})^{2} do = \frac{4 \pi}{9} \alpha a^{2}_{0} \; \Bigl(\frac{p}{\omega}\Bigr) \; \sin^{2}\Theta \mid I_{rd} \mid^{2} do \; \; 
 \label{sigma-aa}
\end{eqnarray} 
where $\Theta$ is the angle between two unit vectors ${\bf n}_{l}$ and ${\bf n}$. The presence of vector product $({\bf n}_{l} \times {\bf 
n})^{2}$ in Eq.(\ref{sigma-aa}) is typical for the dipole approximation. As follows from the formula, Eq.(\ref{sigma-aa}), analytical and 
numerical calculations of the differential cross section of photodetachment are now reduced to analytical computations of the auxiliary 
radial integral $I_{rd}$, Eq.(\ref{aux}). The total cross section of photodetachment is 
\begin{eqnarray}
 \sigma = \Bigl(\frac{32 \pi^{2} \alpha a^{2}_{0} p}{27 \omega}\Bigr) \; \mid I_{rd} \mid^{2} = \frac{32 \pi \alpha a^{2}_{0}}{27} \; 
 \Bigl(\frac{p}{\omega}\Bigr) \mid I_{rd} \mid^{2} \; \; \label{sigma-aa1}
\end{eqnarray} 
By using different expressions for the initial and final wave functions we can determine the differential and total cross sections of 
photodetachment of the outer most electrons in various few- and many-electron atoms and ions. The corresponding formulas are presented 
below. 

\section{Photodetachment of the outer-most electron(s) in few-electron neutral atoms and positively charged ions} 

First, let us consider photodetachment of the outer-most electron(s) in few-elecron neutral atoms, where $Q = N_e$ and $N_e \ge 2$, and in 
positively charged atomic ions, where $Q > N_e$ and $N_e \ge 2$. In both these cases the final sub-systems, i.e., outgoing photo-electron 
and remaining positively charged ion, interact with each other by an attractive Coulomb potential. For atoms and positively charged ions 
this process obviously coincides with atomic photoionization. The wave function of outgoing photoelectron must be taken in the form of 
Eq.(\ref{rad1}), while the wave function of the initial atomic state is chosen in the form of Eq.(\ref{rad3}). The radial derivative of this 
initial wave function is
\begin{eqnarray} 
  \frac{d}{d r} \Bigl[ r^{b - 1} \exp(- B r) \Bigr] = (b - 1) r^{b - 2} \exp(- B r) - B r^{b - 1} \exp(- B r) \; \; \label{diff1} 
\end{eqnarray}  
where $b = \frac{Q - N_e + 1}{\sqrt{2 I}} = \frac{Z}{\sqrt{2 I}} = \frac{Z}{B}, Z = Q - N_e + 1$ and $B = \sqrt{2 I}$. Therefore, the 
formula for our auxiliary radial integral $I_{rd}$, Eq.(\ref{aux}), includes the two terms $I_{rd} = I^{(1)}_{rd} + I^{(2)}_{rd}$, where 
\begin{eqnarray} 
 & &I^{(1)}_{rd} = p \sqrt{\frac{2 \pi \nu (1 + \nu^{2})}{1 - \exp(-2 \pi \nu)}} C(b;B) (b - 1)
 \int_{0}^{+\infty} r^{(b + 2) - 1} \exp(- B r -\imath p r) {}_1F_{1}(2 + \imath \nu; 4; \nonumber \\
 & &2 \imath p r) dr = p \sqrt{\frac{2 \pi \nu (1 + \nu^{2})}{1 - \exp(-2 \pi \nu)}} C(b;B) \; \frac{(b - 1) \Gamma(b + 2)}{(B 
 + \imath p)^{b + 2}} \; \; {}_2F_{1}(2 + \imath \nu; b + 2; 4; \frac{2 \imath p}{B + \imath p} \Bigr) \; \; \label{rint1} \\
 &=&p \sqrt{\frac{2 \pi \nu (1 + \nu^{2})}{1 - \exp(-2 \pi \nu)}} C(b;B) \; \frac{(b - 1) \Gamma(b + 2)}{(B + \imath p)^{b + 2}} 
 \Bigl(\frac{B + \imath p}{B - \imath p}\Bigr)^{\imath \nu + b} \; \; {}_2F_{1}(2 - \imath \nu; 2 - b; 4; \frac{B - \imath p}{B 
 + \imath p} \Bigr) \nonumber  
\end{eqnarray}  
where ${}_2F_{1}(a, b ; c; z)$ is the (2,1)-hypergeometric function defined exactly as in \cite{GR}, $\nu = \frac{Z}{p}$ and $Z = 
Q - N_e + 1$, while $C(b;B)$ is the normalization constant of the bound state radial function (see,  Eq.(\ref{norm})). The explicit 
formula for the second radial integral $I^{(2)}_{rd}$ is 
\begin{eqnarray} 
 & &I^{(2)}_{rd} = - p \sqrt{\frac{2 \pi \nu (1 + \nu^{2})}{1 - \exp(-2 \pi \nu)}} C(b;B) \; B \int_{0}^{+\infty} r^{(b + 3) - 1} 
 \exp(- B r -\imath p r) {}_1F_{1}(2 + \imath \nu; 4; \nonumber \\
 &2& \imath p r) dr = - p \sqrt{\frac{2 \pi \nu (1 + \nu^{2})}{1 - \exp(-2 \pi \nu)}} C(b;B) \; \frac{B \; \Gamma(b + 3)}{(B + 
 \imath p)^{b + 3}} \; \; {}_2F_{1}(2 + \imath \nu; b + 3; 4; \frac{2 \imath p}{B + \imath p} \Bigr) \; \label{rint2} \\
 &=& -p \sqrt{\frac{2 \pi \nu (1 + \nu^{2})}{1 - \exp(-2 \pi \nu)}} C(b;B) \; \frac{B \; \Gamma(b + 3)}{(B + \imath p)^{b + 3}} 
 \Bigl(\frac{B + \imath p}{B - \imath p}\Bigr)^{\imath \nu + b + 1} \; \; {}_2F_{1}(2 - \imath \nu, 1 - b; 4; \frac{B - 
 \imath p}{B + \imath p} \Bigr) \nonumber  
\end{eqnarray} 
where $C(b;B)$ is the normalization constant, Eq.(\ref{norm}), and $B = \sqrt{2 I}$, where $I$ is the ionization potential of the 
initial atomic systems. To simplify the two last formulas we note that 
\begin{eqnarray} 
 \Bigl(\frac{B + \imath p}{B - \imath p}\Bigr)^{\imath \nu} = \Bigl[ \Bigl(\frac{\frac{\nu}{b} + \imath}{\frac{\nu}{b} - 
 \imath}\Bigr)^{\imath \frac{\nu}{b}} \Bigr]^{b} = \exp\Bigl[- 2 \nu \; {\rm arccot} \Bigl(\frac{\nu}{b}\Bigr)\Bigr] \; \label{artg}
\end{eqnarray} 
where $\nu = \frac{Z}{p} = \frac{Q - N_e + 1}{p}$ and $b = \frac{Q - N_e + 1}{\sqrt{2 I}}$ (here $Z = Q - N_e + 1$, see above) and 
$\frac{B}{p} = \frac{\nu}{b}$, or $\nu = \frac{b B}{p}$. The function ${\rm arccot} \; x$ is the inverse cotangent function which 
is also equals ${\rm arccot} \; x = \arccos (\frac{x}{\sqrt{1 + x^{2}}})$ (this formula is often used in numerical calculations). 

After a few additional, relatively simple transformations we arive to the following expression for the total radial integral $I_{rd} 
= I^{(1)}_{rd} + I^{(2)}_{rd}$:
\begin{eqnarray} 
 &&I_{rd} = \frac{p^{2}}{b} \sqrt{\frac{\nu (1 + \nu^{2})}{1 - \exp(-2 \pi \nu)}} \; \frac{2^{b} B^{b} \sqrt{B} \Gamma(b + 2) (B - 
 \imath p)^{1 - b}}{\sqrt{\Gamma(2 b + 1)} (B^2 + p^{2})^{2}} \exp\Bigl[- 2 \nu \; {\rm arccot} \Bigl(\frac{\nu}{b}\Bigr)\Bigr] \Bigr[ 
 (b - 1) \times \nonumber \\
 &&(\nu - \imath b) \; \; {}_2F_{1}(2 - \imath \nu; 2 - b; 4; \frac{\nu - \imath b}{\nu + \imath b} \Bigr) - \nu \; (b + 2) \; \; 
 {}_2F_{1}(2 - \imath \nu, 1 - b; 4; \frac{\nu - \imath b}{\nu + \imath b} \Bigr) \Bigr]  \; \label{Ird}
\end{eqnarray}
From this expression one easily finds the following formula for the $\mid I_{rd} \mid^{2}$ value 
\begin{eqnarray} 
 &&\mid I_{rd} \mid^{2} = \frac{4^{b} \; \nu^{2 b + 2} (1 + \nu^{2}) \Gamma^{2}(b + 2)}{p (\nu^{2} + b^{2})^{b+3} (1 - \exp(-2 \pi 
 \nu)) \Gamma(2 b + 1)} \; \exp\Bigl[- 4 \nu \; {\rm arccot} \Bigl(\frac{\nu}{b}\Bigr)\Bigr] \; \Bigr| (b - 1) \times \nonumber \\
 && (\nu - \imath b) \; \; {}_2F_{1}(2 - \imath \nu; 2 - b; 4; \frac{\nu - \imath b}{\nu + \imath b} \Bigr) - \nu \; (b + 2) \; 
 {}_2F_{1}(2 - \imath \nu, 1 - b; 4; \frac{\nu - \imath b}{\nu + \imath b} \Bigr) \Bigr|^{2} \; \label{I2rdmod}
\end{eqnarray}
By multiplying this expression by the $\Bigl(\frac{8 \pi \alpha a^{2}_{0} p}{9 \omega}\Bigr) \; ({\bf n} \times {\bf e})^{2}$ factor 
one finds the final formula for the differential cross section of photoionization of few- and many-electron neutral atoms and/or 
positively charged ions each of which contains $N_e$ bound electrons ($N_e \ge 2$) and one central atomic nucleus with the electrical 
charge $Q$ (or $Q e$)  
\begin{eqnarray} 
 &&d\sigma = \Bigl(\frac{8 \pi \alpha a^{2}_{0}}{9 \omega}\Bigr) \; \frac{4^{b} \; \nu^{2 b + 2} (1 + \nu^{2}) \Gamma^{2}(b + 
 2)}{(\nu^{2} + b^{2})^{b+3} (1 - \exp(-2 \pi \nu)) \Gamma(2 b + 1)} \; \exp\Bigl[- 4 \nu \; {\rm arccot}\Bigl(\frac{\nu}{b}\Bigr)\Bigr] 
 \Bigr| (b - 1) \times \nonumber \\
 && (\nu - \imath b) \; \; {}_2F_{1}(2 - \imath \nu; 2 - b; 4; \frac{\nu - \imath b}{\nu + \imath b} \Bigr) - \nu \; 
 (b + 2) \; {}_2F_{1}(2 - \imath \nu, 1 - b; 4; \frac{\nu - \imath b}{\nu + \imath b} \Bigr) \Bigr|^{2} \; \label{dsigmaZpol} \\
 && ({\bf n} \cdot {\bf e})^{2} do \; \nonumber 
\end{eqnarray} 
where the incident beam of light is completely polarized. For natural light we obtain the following formula 
\begin{eqnarray} 
 &&d\sigma = \Bigl(\frac{4 \pi \alpha a^{2}_{0}}{9 \omega}\Bigr) \; \frac{4^{b} \; \nu^{2 b + 2} (1 + \nu^{2}) \Gamma^{2}(b + 
 2)}{(\nu^{2} + b^{2})^{b+3} (1 - \exp(-2 \pi \nu)) \Gamma(2 b + 1)} \; \exp\Bigl[- 4 \nu \; {\rm arccot} \Bigl(\frac{\nu}{b}\Bigr)\Bigr] 
 \Bigr| (b - 1) \times \nonumber \\
 && (\nu - \imath b) \; \; {}_2F_{1}(2 - \imath \nu; 2 - b; 4; \frac{\nu - \imath b}{\nu + \imath b} \Bigr) - \nu \; (b + 2) \; 
 {}_2F_{1}(2 - \imath \nu, 1 - b; 4; \frac{\nu - \imath b}{\nu + \imath b} \Bigr) \Bigr|^{2} \times \; \label{dsigmaZu} \\
 && ({\bf n}_l \times {\bf n})^{2} do \; \nonumber 
\end{eqnarray}
 
Finally, the corresponding formula for the total cross section of photoionization of the $(Q, N_e)-$atomic system, where $N_e \ge 2$,
takes the form
\begin{eqnarray} 
 && \sigma = \Bigl(\frac{32 \pi^{2} \alpha a^{2}_{0}}{27 \omega}\Bigr) \; \frac{4^{b} \; \nu^{2 b + 2} (1 + \nu^{2}) \Gamma^{2}(b + 
 2)}{(\nu^{2} + b^{2})^{b+3} (1 - \exp(-2 \pi \nu)) \Gamma(2 b + 1)} \; \exp\Bigl[- 4 \nu \; {\rm arccot}\Bigl(\frac{\nu}{b}\Bigr)\Bigr] 
 \times \; \label{sigmaZ} \\
 && \Bigr| (b - 1) (\nu - \imath b) \; \; {}_2F_{1}(2 - \imath \nu; 2 - b; 4; \frac{\nu - \imath b}{\nu + \imath b} \Bigr) - \nu \; 
 (b + 2) \; {}_2F_{1}(2 - \imath \nu, 1 - b; 4; \frac{\nu - \imath b}{\nu + \imath b} \Bigr) \Bigr|^{2} \; \nonumber
\end{eqnarray} 
Note that the parameter $b$ in these formulas is a real number, which is usually bounded between 0 and 2, i.e., $0 < b < 2$. This means 
that the hypergeometric functions in Eqs.(\ref{rint1}) and (\ref{rint2}) can be determined only numerically (two exceptional cases when 
$b = 1$ and $b = 0$ are considered in the next two Sections). Recently, a number of fast, reliable and numerically stable algorithms 
have been developed and tested for accurate calculations of the hypergeometric functions. Our final formula can be simplified even 
further, if one applies the following relation (see, e.g., Eq.15.2.3 in \cite{AS}) between two hypergeometric functions which are 
included in our Eq.(\ref{I2rdmod}):
\begin{eqnarray} 
 & &{}_2F_{1}(2 - \imath \nu, 2 - b; 4; \frac{\nu - \imath b}{\nu + \imath b} \Bigr) = 
  {}_2F_{1}(2 - \imath \nu, 1 - b; 4; \frac{\nu - \imath b}{\nu + \imath b} \Bigr) \nonumber \\
 &+& \frac{1}{1 - b} \Bigl(\frac{\nu - \imath b}{\nu + \imath b}\Bigr) \frac{d }{d z} \Bigl[ {}_2F_{1}(2 - \imath \nu, 
 1 - b; 4; z) \Bigr] \; \; \label{15.2.3}
\end{eqnarray}
where $z = \frac{\nu - \imath b}{\nu + \imath b}$. This formula allows one to operate with one hypergeometric function only. 

All formulas derived and presented in this Section can directly be used to determine the both differential and total cross sections of 
photodetachment of the outer-most electrons in few- and many-electron neutral atoms and positively charged ions which contains $N_e$ bound 
electron, where $N_e \ge 2$. The notation $Q = Q e$ is the electrical charge of the central atomic nucleus (in atomic units). For 
one-electron atoms and positively charged ions our formulas derived above simplified significantly. Photoionization of one-electron atoms 
and ions, in which $N_e = 1$, is considered in the next Section. 

To conlude this Section, we want to note that the total cross section of photoionization of the $(Q, N_e)-$atomic system, Eq.(\ref{sigmaZ}) 
can be re-written in a slightly different form $\sigma(\omega) = \sigma(\omega; I, b)$ which shows the explicit dependence of the 
photoionization cross section upon the cyclic frequency of incident light. Such explicit formulas are very popular among experimenters and 
those theorists who calculate the convolution of various energy spectra, e.g., thermal spectra, and photodetachment cross sections. To 
achive this goal one needs to replace variables in Eq.(\ref{sigmaZ}) by using the following (equivalent) expression for $\nu$ written in 
terms of $I, \omega$ and $b$: $\nu = b \sqrt{\frac{I}{\omega - I}}$. Finally, one finds the following formula for the total cross section 
$\sigma = \sigma(\omega, I, b)$ of photodetachment of the outer-most electrons in few-electron neutral atoms and positively charged ions
\begin{eqnarray} 
\sigma&=&\Bigl(\frac{32 \pi^{2} \alpha a^{2}_{0}}{27}\Bigr) \; \frac{4^{b} \; I^{b + 2} \; \Bigl(1 + (b^{2} - 1) \frac{I}{\omega}\Bigr) 
 \; \; \Gamma^{2}(b + 3)}{b^{4} \omega^{b+3} \; [1 - \exp(-2 \pi b \sqrt{\frac{I}{\omega - I}}\Bigr)\Bigr] \; \Gamma(2 b + 1)} \; 
 \exp\Bigl[- 4 b \sqrt{\frac{I}{\omega - I}} \; {\rm arccot}\Bigl(\sqrt{\frac{I}{\omega - I}}\Bigr)\Bigr] \nonumber \\
 &\times&\Bigl| \frac{b - 1}{b + 2} \; \Bigl(1 - \imath b \sqrt{\frac{\omega - I}{I}}\Bigr) \; \; {}_2F_{1}(2 - \imath b 
 \sqrt{\frac{I}{\omega - I}}, 2 - b; 4; \frac{\sqrt{I} - \imath \sqrt{\omega - I}}{\sqrt{I} + \imath \sqrt{\omega - I}} \Bigr) \nonumber \\
 &-& {}_2F_{1}(2 - \imath b \sqrt{\frac{I}{\omega - I}}, 1 - b; 4; \frac{\sqrt{I} - \imath \sqrt{\omega - I}}{\sqrt{I} 
 + \imath \sqrt{\omega - I}} \Bigr) \Bigr|^{2} \; \label{sigmaZ1}
\end{eqnarray} 
This formula is one of the main results of this study. For one-electron stomic systems, when $b = 1$, one finds directly from the last formula 
\begin{eqnarray} 
\sigma&=&\Bigl(\frac{32 \pi^{2} \alpha a^{2}_{0}}{27}\Bigr) \; \frac{4 \; I^{3} \; \; \Gamma^{2}(4)}{\omega^{4} \; [1 - \exp(-2 \pi  
 \sqrt{\frac{I}{\omega - I}}\Bigr)\Bigr] \; \Gamma(3)} \; \exp\Bigl[- 4 \sqrt{\frac{I}{\omega - I}} \; {\rm arccot} \; 
 \Bigl(\sqrt{\frac{I}{\omega - I}}\Bigr)\Bigr] \nonumber \\ 
 &=& \frac{256 \pi^{2} \alpha a^{2}_{0}}{3} \; \frac{I^{3}}{\omega^{4} \; [1 - \exp(-2 \pi \sqrt{\frac{I}{\omega - I}}\Bigr)\Bigr]} \; 
 \exp\Bigl[- 4 \sqrt{\frac{I}{\omega - I}} \; {\rm arccot} \; \Bigl(\sqrt{\frac{I}{\omega - I}}\Bigr)\Bigr] \; \; \label{FFF}
\end{eqnarray} 
i.e., the formula which exactly coincides with our formula, Eq.(\ref{sigmaQ}), derived below, since for $b = 1$ (or for one-electron atomic 
systems) we have $I = \frac{Q^{2}}{2}$ in atomic units.   

\section{Photoionization of one-electron atoms and ions}

Photoionization of one-electron atomic systems is significantly simpler than photodetachment of the outer-most electrons in a few-and many-electron 
atomic systems considered above. Indeed, in this case we have the atomic ionization potential $I$ which depends upon the nuclear charge $Q$ only, 
i.e., $I = I(Q)$. Furthermore, for the ground (bound) state in one-electron atoms and ions we always have $2 I = Q^{2}$, and therefore, in the both 
formulas, Eqs.(\ref{rint1}) and Eq.(\ref{rint2}) the parameter $b = 1$ and the first term in Eq.(\ref{diff1}) equals zero identically. Also, for $b 
= 1$ the hypergeometric functions ${}_2F_{1}(2 - \imath \nu, 0; 4; z)$, which is included in the second term, equals unity. This means that for 
one-electron atoms/ions (or for $b = 1$) we can express the both differential and total cross sections of photoionization in terms of elementary 
functions only. The normalization constant of the incident wave function equals $C(b;B) = C(1;Q) = \frac{Q \sqrt{Q}}{\sqrt{\pi}}$ and the two 
paramaters $\nu$ and $\frac{\nu}{b}$ are now identical. Briefly, we can say that for any one-electron atom and/or ion the long-range asymptotics of 
its actual wave function always coincides with the wave function itself, and for ground states it is also concides with the formula, Eq.(\ref{rad3}). 
In other words, by applying our method to the ground states in one-electron atomic systems one obtains the complete and exact solution for the problem 
of non-relativistic photoionization. 

For one-electron atoms and ions the explicit formula for the auxiliary radial integral $I_{rd}$ in Eq.(\ref{ME}) (or $I^{(2)}_{rd}$ in 
Eq.(\ref{diff1})) takes the form    
\begin{eqnarray} 
 & &I_{rd} = - p Q^{2} \frac{Q \sqrt{Q}}{\sqrt \pi} \sqrt{\frac{2 \pi (1 + \nu^{2})}{\nu (1 - \exp(-2 \pi \nu))}} \; 
 \int_{0}^{+\infty} r^{(4 - 1)} \exp(- Q r -\imath p r) {}_1F_{1}(2 + \imath \nu; 4; 2 \imath p r) dr \nonumber \\
 &=& - p Q^{3} \sqrt{Q} \sqrt{\frac{2 (1 + \nu^{2})}{\nu (1 - \exp(-2 \pi \nu))}} \; \frac{\Gamma(4)}{(Q + \imath p)^{4}} \; \; 
 {}_2F_{1}(2 + \imath \nu; 4; 4; \frac{2 \imath p}{Q + \imath p} \Bigr) \; \label{rint2a}
\end{eqnarray} 
where $\nu = \frac{Q}{p}$ and $\Gamma(4) = 3 \cdot 2 \cdot 1 = 6$. The hypergeometric function can be transformed to the form 
\begin{eqnarray}
 {}_2F_{1}(2 + \imath \nu; 4; 4; \frac{2 \imath p}{Q + \imath p} \Bigr) = \Bigl(1 - \frac{2 \imath p}{Q + \imath p} 
 \Bigr)^{4 - 4 - 2 - \imath \nu} {}_2F_{1}(2 - \imath \nu; 0; 4; \frac{2 \imath p}{Q + \imath p} \Bigr) = \Bigl(\frac{Q + 
 \imath p}{Q - \imath p} \Bigr)^{2 + \imath \nu} \; \label{hgf}
\end{eqnarray} 
by using the formula ${}_2F_{1}(\alpha, \beta; \gamma; z) = (1 - z)^{\gamma - \alpha - \beta} {}_2F_{1}(\gamma - \alpha, \gamma - \beta; 
\gamma; z)$ (see, Eq.(9.131) in \cite{GR}). Another way to obtain the same formula, Eq.(\ref{hgf}), is to apply the following expression 
for the integral in Eq.(\ref{rint2a}) with the confluent hypergeometric function(s) (see, e.g., \cite{LLQ} and \cite{GR})
\begin{eqnarray}
 \int_{0}^{+\infty} \exp(- \lambda z) z^{\gamma - 1} {}_1F_{1}(\alpha; \gamma; b z) dz = \frac{\Gamma(\gamma)}{\lambda^{\gamma}} \;
 \Bigl(\frac{\lambda}{\lambda - b}\Bigr)^{\alpha}
\end{eqnarray}
Finally, we obtain 
\begin{eqnarray} 
 & &I_{rd} = - 6 Q^{3} \sqrt{Q} \sqrt{\frac{2 (1 + \nu^{2})}{\nu (1 - \exp(-2 \pi \nu))}} \; \frac{1}{(Q^{2} + p^{2})^{2}}  
 \; \Bigl(\frac{\nu + \imath}{\nu - \imath}\Bigr)^{\imath \nu} \nonumber \\
 &=& - \frac{6 Q^{3} \sqrt{Q}}{p^{4}} \sqrt{\frac{2}{\nu (1 - \exp(-2 \pi \nu)) \; (1 + \nu^{2})^{3}}} \; \frac{1}{(1 + 
 \nu^{2})^{2}} \; \exp( -2 \nu \; {\rm arccot} \; \nu) \label{1Ird}
\end{eqnarray} 

By using the equatily $\nu = \frac{Q}{p}$ and multiplying the radial integral $I_{rd}$ by the factor $\frac{4 \pi}{3} ({\bf n} \cdot 
{\bf e})$ one obtains the following formula 
\begin{eqnarray} 
 \frac{4 \pi}{3} ({\bf n} \cdot {\bf e}) I_{rd} = - \frac{8 \pi}{\sqrt{p}} \sqrt{\frac{2 \nu^{6}}{[1 - \exp(-2 \pi \nu)] 
 (1 + \nu^{2})^{3}}} \; \exp( -2 \nu \; {\rm arccot} \; \nu) \label{1Irda}
\end{eqnarray} 
The explicit formula for the matrix element $\frac{p \alpha a^{2}_{0}}{2 \pi \omega} \mid I_{rd} \mid^{2}$ factor takes the form 
\begin{eqnarray} 
 \frac{p \alpha a^{2}_{0}}{2 \pi \omega} \mid I_{rd} \mid^{2} = \frac{64 \pi \alpha a^{2}_{0}}{\omega} \Bigl(\frac{\nu^{2}}{1 + 
 \nu^{2}}\Bigr)^{3} \; \frac{\exp( -4 \nu \; {\rm arccot} \; \nu)}{1 - \exp(-2 \pi \nu)} \label{MatrEl}
\end{eqnarray} 
Theferore, the final formula for the differential cross sections of photoionization of one-electron atom/ion by completely polarized 
light is written in the form 
\begin{eqnarray} 
 d\sigma &=& \frac{64 \pi \alpha a^{2}_{0}}{\omega} \Bigl(\frac{\nu^{2}}{1 + \nu^{2}}\Bigr)^{3} \; \frac{\exp( -4 \nu \; {\rm arccot} 
 \; \nu)}{1 - \exp(-2 \pi \nu)} ({\bf n} \cdot {\bf e})^{2} do \nonumber \\
 &=& \frac{128 \pi \alpha a^{2}_{0}}{Q^{2}} \Bigl(\frac{I}{\omega}\Bigr)^{4} \; \frac{\exp( -4 \nu \; {\rm arccot} \; \nu)}{1 - 
 \exp(-2 \pi \nu)} ({\bf n} \cdot {\bf e})^{2} do \; \; \label{dsigmaQpol}
\end{eqnarray} 
where we also used the following relation $1 = \frac{Q^{2}}{Q^{2}} = 2 \Bigl(\frac{Q^{2}}{2}\Bigr) \frac{1}{Q^{2}} = \frac{2 I}{Q^{2}}$. 
For the natural (or unpolarized) light the differential cross section of photoionization is written in the form
\begin{eqnarray} 
 d\sigma =  \frac{64 \pi \alpha a^{2}_{0}}{Q^{2}} \Bigl(\frac{I}{\omega}\Bigr)^{4} \; \frac{\exp( -4 \nu \; {\rm arccot} \; \nu)}{1 - 
 \exp(-2 \pi \nu)} ({\bf n} \times {\bf n}_{l})^{2} do \; \; \label{dsigmauQunpol} 
\end{eqnarray}
The total cross section of photoionization for one-electron atom/ion with the nuclear charge $Q$ is 
\begin{eqnarray}
 \sigma &=& 512 \pi^{2} \alpha \; \Bigl(\frac{a^{2}_{0}}{Q^{2}}\Bigr) \; \Bigl(\frac{I}{\omega}\Bigr)^{4} \; \frac{\exp( -4 \nu \;  
 {\rm arccot} \; \nu)}{1 - \exp(-2 \pi \nu)} \nonumber \\
 &=& 512 \pi^{2} \alpha \; \Bigl(\frac{a^{2}_{0}}{Q^{2}}\Bigr) \; \Bigl(\frac{I}{\omega}\Bigr)^{4} \; 
 \frac{\exp\Bigl( -4 \sqrt{\frac{I}{\omega - I}} \; {\rm arccot} \; \sqrt{\frac{I}{\omega - I}}\Bigr)}{1 - \exp\Bigl(-2 \pi 
 \sqrt{\frac{I}{\omega - I}}\Bigr)} \;  \; \; \label{sigmaQ} 
\end{eqnarray} 
This result coincides with the formula obtained earlier by Stobbe in 1930 \cite{Stob}. Our formulas, Eqs.(\ref{dsigmaQpol}) - (\ref{sigmaQ}), 
exactly coincides with the analogous formulas derived in \$ 57 in \cite{BLP} and \$ 34 in \cite{Sobelman}. Note that the last formulas, 
Eqs.(\ref{dsigmaQpol}) - (\ref{sigmaQ}), can be derived from our formulas presented in the previous Section for $b = 1$, but here we wanted to 
derive them by using an independent approach. As mentioned above for each one-electron atoms and ions its ionization potential $I$ is the 
explicit (and simple) function of the nuclear charge $Q$ only. Generalization of our formulas to photodetachment of the outer-most electron 
from the excited atomic states is also simple and transparent, but it cannot be done directly with the use of DFT theory, since Eq.(\ref{rad3}) 
is wrong for the excited states. Derivtion of the explicit formulas for the photoionization cross sections also requires some additional work, 
notations and explanations (some details of such calculations are discissed in the Appendix).  

\section{Photodetachment of the negatively charged ions} 

Photodetachment of the negatively charged atomic ions is of great interest in numerous applications (see, the first Section (Introduction) above). 
In general, there is a fundamental difference in photodetachment of the negatively charged ions and positively charged atomic ions and neutral 
atoms. In particular, for all negatively charged ions their effective electrical charge $Z = Q - N_e + 1$ equals zero identically. Therefore, in 
this case we cannot introduce the parameter $\nu = \frac{Z}{p}$, which was used in the two previous Sections. This means that all our formulas, 
derived for the photodetachment cross sections (see below), will contain only the momentum of photoelectron $p$ and ionization potential $I$, or 
parameter $B = \sqrt{2 I}$ which is extensively used below. These two variables $p$ and $I$ (or $B$) are crucial for theoretical analysis of the 
non-relativistic photodetachment of the negatively charged ions. As mentioned above photodetachment of the negatively charged, two-electron 
H$^{-}$ ion is of interest for understanding of the actual visible and infrared spectra of some stars, including our Sun (see discussion and 
references in \cite{Fro2015A}). Analogous process for the negatively charged Ps$^{-}$ ion was considered in \cite{BD1} and \cite{Fro2015}. 
Photodetachment of the four-electron negatively charged Li$^{-}$ ion plays some role in developing of very compact and reliable photoelements 
and recharged batteries. Therefore, it is important to produce the general formula for the photodetachment cross sections of the negatively 
charged ions.    

For the negatively charged atomic ions the corresponding derivative of the radial wave function of the initial state is written as follows 
\begin{eqnarray} 
  \frac{d}{d r} \Bigl[ \frac{C}{r} \exp(- B r) \Bigr] = - C r^{-2} \exp(- B r) - C B r^{-1} \exp(- B r) \; \; \label{diff2} 
\end{eqnarray} 
where $C$ is the normalization constant which equals $\sqrt[4]{\frac{I}{2 \pi^{2}}}$ as this follows from Eq.(\ref{norm}) for $b = 0$.
Therefore, the formula for our auxiliary radial integral $I_{rd}$, Eq.(\ref{aux}), also includes two terms, i.e., $I_{rd} = 3 \Bigl( J^{(1)}_{rd} 
+ J^{(2)}_{rd} \Bigr)$, where 
\begin{eqnarray} 
 & &J^{(1)}_{rd} = C \sqrt{\frac{\pi}{2 p}} \int_{0}^{+\infty} dr J_{\frac32}(p r) r^{\frac12 - 1} \exp(- B r) = C \sqrt{\frac{\pi}{2 p}} 
 \Bigl(\frac{p}{2}\Bigr)^{\frac32} \frac{\Gamma(2)}{\Gamma\Bigl(\frac52\Bigr) (B^{2} + p^{2})} \times \; \; \label{J1a} \\
 & &{}_2F_{1}\Bigl(1, 1; \frac52; \frac{p^{2}}{B^{2} + p^{2}} \Bigr) = \frac{C p}{3 (B^{2} + p^{2})} \; \; {}_2F_{1}\Bigl(\frac32, \frac32; 
 \frac52; \frac{p^{2}}{B^{2} + p^{2}} \Bigr) \nonumber
\end{eqnarray} 
The hypergeometric function in the last equation can be reduced to some combination of elementary functions. To show this explicitly let us 
apply the following formula 
\begin{eqnarray}
 \frac{d}{d z} \; \; \Bigl[ {}_2F_{1}(\alpha, \beta; \gamma; z) \Bigr] = \frac{\alpha \beta}{\gamma} \; \; {}_2F_{1}(\alpha + 1, \beta + 1; 
\gamma + 1; z) \; \; 
 \label{HHG1}
\end{eqnarray} 
where in our case $\alpha = \frac12, \beta = \frac12$ and $\gamma = \frac32$. For these values of $\alpha, \beta$ and $\gamma$ the last formula 
takes the form 
\begin{eqnarray}
 \frac{d}{d z} \; \; \Bigl[ {}_2F_{1}\Bigl( \frac12, \frac12; \frac32; z ) \Bigr] = \frac16 \; \; {}_2F_{1}\Bigl( \frac32, \frac32;  \frac52; 
 z \Bigr) \; \; \label{HHG2}
\end{eqnarray} 
where the argument $z$ varies between zero and unity, i.e., $0 \le z \le 1$. In our case this is true, since $z = \frac{p^{2}}{B^{2} + p^{2}}$. 
Now, we can write 
\begin{eqnarray}
 {}_2F_{1}\Bigl( \frac32, \frac32; \frac52; z\Bigr) = 6 \frac{d}{d z} \Bigl[ {}_2F_{1}\Bigl(\frac12; \frac12; \frac32; z) \Bigr] = 6 
 \frac{d}{d z} \Bigl( \frac{\arcsin \sqrt{z}}{\sqrt{z}} \Bigr) = 3 \frac{\sqrt{z} - \sqrt{1 - z} \; \arcsin \sqrt{z}}{z \sqrt{z 
 (1 - z)}} \; \; . \label{HHG3}
\end{eqnarray} 
where we used the formula Eq.(15.1.6) from \cite{AS} for the ${}_2F_{1}(\frac12; \frac12; \frac32; z)$ function, i.e., ${}_2F_{1}(\frac12; 
\frac12; \frac32; z) = \frac{\arcsin \sqrt{z}}{\sqrt{z}}$. The analytical formula, Eq.(\ref{HHG3}), derived for the ${}_2F_{1}\Bigl( 
\frac32, \frac32; \frac52; z \Bigr)$ function is our original result which cannot be found directly neither in \cite{GR}, nor in 
\cite{AS}. In our case $z = \frac{p^{2}}{B^{2} + p^{2}}, \sqrt{1 - z} = \frac{B}{\sqrt{B^{2} + p^{2}}}$ and $\sqrt{z} = \frac{p}{\sqrt{B^{2} 
+ p^{2}}}$ and the final formula for the $J^{(1)}_{rd}$ integral takes the form 
\begin{eqnarray} 
 J^{(1)}_{rd} = \frac{C}{\sqrt{B^{2} + p^{2}}} \; \; \frac{\sqrt{z} - \sqrt{1 - z} \arcsin\sqrt{z}}{z \sqrt{1 - z}} \; \; 
 \label{J1rd}
\end{eqnarray} 
The expression in the right-hand side of this equation is not singular when $z \rightarrow 0$ (or $p \rightarrow 0$), since
\begin{eqnarray} 
 \lim_{z \rightarrow 0} \frac{\sqrt{z} - \sqrt{1 - z} \arcsin\sqrt{z}}{z \sqrt{1 - z}} = \frac16 \lim_{z \rightarrow 0} \sqrt{z} = 0 
 \; \; \nonumber 
\end{eqnarray} 

Analogous formula for the radial integral $J^{(2)}_{rd}$ is 
\begin{eqnarray} 
 & &J^{(2)}_{rd} = C \; B \sqrt{\frac{\pi}{2 p}} \int_{0}^{+\infty} dr J_{\frac32}(p r) r^{\frac32 - 1} \exp(- B r) = \frac{C}{\sqrt{2 p}} 
 \Bigl(\frac{p}{2}\Bigr)^{\frac32} \frac{B \; \Gamma(3)}{\Gamma\Bigl(\frac52\Bigr) (B^{2} + p^{2})^{\frac32}} \times \; \; \nonumber \\
 & &{}_2F_{1}(\frac32; \frac12; \frac52; \frac{p^{2}}{B^{2} + p^{2}} \Bigr) = \frac{2 \; C \; B \; p}{3 (B^{2} + p^{2})^{\frac32}} \; \; 
 {}_2F_{1}(\frac32; \frac12; \frac52; \frac{p^{2}}{B^{2} + p^{2}} \Bigr) \; \; \label{J2a}
\end{eqnarray}
It is possible to obtain the explicit expression of the ${}_2F_{1}\Bigl(\frac32; \frac12; \frac52; z)$ function in terms of some elementary 
functions. For this purpose we need to use the known analytical formula for the ${}_2F_{1}\Bigl( \frac12, \frac12; \frac32; z )$ function
(which equals $\frac{\arcsin \sqrt{z}}{\sqrt{z}}$, see, Eq.(\ref{HHG3})) and apply the following formula Eq.(15.2.7) from \cite{AS} for $n 
= 1$
\begin{eqnarray}
 \frac{d }{d z} \Bigl[ (1 - z)^{a} {}_{2}F_{1}( a, b; c; z) \Bigr] = - \frac{a (c - b)}{c} (1 - z)^{a - 1} {}_{2}F_{1}( a + 1, b; c + 1; z) 
 \label{ASint1}  
\end{eqnarray} 
where $a = \frac12, b = \frac12$ and $c = \frac32$. Now, for the ${}_2F_{1}\Bigl( \frac12, \frac12; \frac32; z )$ function one finds 
\begin{eqnarray}
 \frac{d }{d z} \Bigl[ (1 - z)^{\frac12} {}_{2}F_{1}\Bigl( \frac12, \frac12; \frac32; z) \Bigr] = - \frac{1}{3} (1 - z)^{-\frac12} \; 
 {}_{2}F_{1}( \frac32, \frac12; \frac52; z) \; \;  \label{ASint2}   
\end{eqnarray}
From this equation we derive
\begin{eqnarray}
 & & {}_{2}F_{1}\Bigl( \frac32, \frac12; \frac52; z) = -3 \sqrt{1 - z} \; \frac{d }{d z} \Bigl[ \sqrt{1 - z} \Bigl(\frac{ \arcsin 
 \sqrt{z}}{\sqrt{z}} \Bigr) \Bigr] = \frac32 \Bigl( \frac{\arcsin \sqrt{z}}{\sqrt{z}} \; \nonumber \\
 &-& \frac{\sqrt{1 - z}}{z} + \frac{1 - z}{z \sqrt{z}} \arcsin\sqrt{z} \Bigr) = \frac{3}{2 \sqrt{z}} \Bigl( \frac{\arcsin \sqrt{z}}{z} 
  - \frac{\sqrt{z (1 - z)}}{z} \Bigr) \; \; , \; \label{ASint25}  
\end{eqnarray}
where $z = \frac{p^{2}}{B^{2} + p^{2}} \le 1$ (it is also clear that $z$ is non-negative). This analytical formula for the ${}_2F_{1}\Bigl( 
\frac32, \frac12; \frac52; z \Bigr)$ function is another original result which cannot be found neither in \cite{GR}, nor in \cite{AS}. Note 
also that analytical formula for this integral can also be derived as a partial derivative of the $J^{(1)}_{rd}$ in respect to the parameter 
$B$. Thus, the both auxiliary radial integrals $J^{(1)}_{rd}$ and $J^{(2)}_{rd}$ are expressed in terms of the elementary functions. In 
particular, the final formula for the $J^{(2)}_{rd}$ integral is
\begin{eqnarray} 
 J^{(2)}_{rd} = \frac{C}{\sqrt{B^{2} + p^{2}}} \; \sqrt{1 - z} \; \Bigl[ \frac{\arcsin \sqrt{z}}{z} - \sqrt{\frac{1 - z}{z}}  
 \Bigr]  \; \; \label{J2rd} 
\end{eqnarray}
where $z = \frac{p^{2}}{B^{2} + p^{2}}, \sqrt{z} = \frac{p}{\sqrt{B^{2} + p^{2}}}$ and $\sqrt{1 - z} = \frac{B}{\sqrt{B^{2} + p^{2}}}$. Again, 
by using the formula, Eq.(1.641) from \cite{GR} for the $\arcsin x$ one can easily show that 
\begin{eqnarray} 
 \lim_{z \rightarrow 0} \Bigl(\frac{\arcsin \sqrt{z}}{z} - \sqrt{\frac{1 - z}{z}} \Bigr) = 0 \; \; \nonumber 
\end{eqnarray} 
which means that our formula for the $J^{(2)}_{rd}$ integral is not singular at $z \rightarrow 0$ (or at $p \rightarrow 0$). The formula 
Eq.(\ref{crossau}) for the differential cross section of photodetachment contains an additional factor $p$ in its numerator. Therefore, such 
a cross-section for an arbitrary negatively charged ion always approaches zero when $p \rightarrow 0$. The same is true for the total 
cross sections of photodetachment of the negatively charged ions. In contrast with this, the photodetachment cross sections (differential and 
total) of atomic systems considered in the two previous Sections, i.e., for neutral atoms and positively charged ions, including one-electron 
systems, are always finite. All these features of photodetachment (or photoionization) cross sections are well known from numerous experiments 
(see, e.g., \cite{Star}, \cite{BS} and references therein). 

Analytical computations of the total auxiliary $I_{rd} = 3 \Bigl( J^{(1)}_{rd} + J^{(2)}_{rd} \Bigr)$ integral and the both differential and 
total cross sections is simple and straightforward. The final formula, Eq.(\ref{sigma-aa}), for the differential cross section of photodetachment 
of the negatively charged atomic ions takes the form 
\begin{eqnarray}
 d\sigma = \Bigl(\frac{8 \pi \alpha a^{2}_{0} p}{9 \omega}\Bigr) \; ({\bf n} \cdot {\bf e})^{2} \mid I_{rd} \mid^{2} do = \frac{8 \alpha 
 a^{2}_{0}}{\omega} \; \Bigl[\frac{p B}{\omega (B^{2} + p^{2})}\Bigr] \; \mid J^{(1)}_{rd} + J^{(2)}_{rd} \mid^{2} ({\bf n} \cdot 
 {\bf e})^{2} do \label{sigma-polar}
\end{eqnarray}
for completely polarized light. Analogous formula for unpolarized light is  
\begin{eqnarray}
 d\sigma = \frac{4 \alpha a^{2}_{0}}{\omega} \; \sqrt{\frac{1 - z}{z}} \; \Bigl[(\sqrt{1 - z} - 1) \frac{\arcsin(\sqrt{z})}{\sqrt{z}} 
 + \frac{1}{\sqrt{1 - z}} - 1 + z \Bigr]^{2} ({\bf n}_{l} \times {\bf n})^{2} do \; \; \; \label{sigma-ang} 
\end{eqnarray}
where $z = \frac{p^{2}}{B^{2} + p^{2}}, \sqrt{z} = \frac{p}{\sqrt{B^{2} + p^{2}}}, \; \sqrt{1 - z} = \frac{B}{\sqrt{B^{2} + p^{2}}}$ and $B^{2} 
= 2 I$. This formula for the differential cross section of photodetachment of the negatively charged atomic ions is one of the main results of 
this study. The formula for the total cross section can be finalized to the form
\begin{eqnarray}
  \sigma &=& \frac{32 \pi \alpha a^{2}_{0}}{3 \omega} \; \sqrt{\frac{1 - z}{z}} \; \Bigl[(\sqrt{1 - z} - 1) \frac{\arcsin(\sqrt{z})}{\sqrt{z}} 
 + \frac{1}{\sqrt{1 - z}} - 1 + z \Bigr]^{2} \; \; \; \nonumber \\
 &=& \frac{32 \pi \alpha a^{2}_{0}}{3 \omega} \; \sqrt{\frac{I}{\omega - I}} \; \Bigl[ \frac{\sqrt{I} - \sqrt{\omega}}{\sqrt{\omega - I}} 
 \arcsin\sqrt{1 - \frac{I}{\omega}} + \sqrt{\frac{\omega}{I}} - \frac{I}{\omega} \Bigr]^{2} \; \; \; \label{sigma-totl} 
\end{eqnarray} 
where $z = 1 - \frac{I}{\omega}$ and $1 - z = \frac{I}{\omega}$. These exact formulas completely solve the problem of photodetachment of the 
outer-most electron in the negatively charged atomic ions. Note also that the formula, Eq.(\ref{sigma-totl}), for the $\sigma = \sigma(\omega, 
I)$ dependence allows one to check and mainly confirm earlier predictions made by Chandrasekhar in his papers about photodetachment 
cross-section of the negatively charged H$^{-}$ ion \cite{Chan1}, \cite{Chan2} (see also discussion in Sect.74 of \cite{BS}, \cite{John} and 
\cite{FroF} and references therein). The formulas derived above can also be used to describe photodetchment of the weakly-bound deuterium nucleus 
\cite{BetPei} (see, also \cite{Bet2}). 

In general, the formula Eq.(\ref{sigma-totl}) well describes photodetachment of the negatively charged ions. In particular, this formula correctly 
represents the qualitative dependence of the $\sigma(\omega)$ function. However, the observed quantitative agreement is not so good and its maximal 
error is on the order of 16 \%. Such an agreement can substantially be improved by entering one (or a few) additional parameters as it was done in 
\cite{Bet2}. In the first approximation we need to multiply our cross sections computed for different frequensies $\omega$ by the numerical factor 
$\lambda$ which equals to the ratio of maximal experimental cross section to the maximum cross section computed with the use of Eq.(\ref{sigma-totl}). 
In our calculations of the photodetachment cross section of the H$^{-}$ ion (results are presented in Table I) we have used the numerical factor 
$\lambda$ = 0.84750. Note that this factor is always less than unity. This follows from the fact that in our method we have replaced the real atomic 
wave function (outside of its asymptotic region) by its asymptotics, and this asymptotics (more precisely, its radial derivative) exceeds the real 
radial wave function (its radial derivative). The same method is applied to the photoionization of the neutral atoms and positively charged ions 
which is considered in Section IV.   

\section{Variational approach for the photoionization and photodetachment cross sections}

All formulas derived above are correct and exact. However, the overall accuracies of these formulas in actual numerical computations are quite 
restricted, since in each of these formulas only one function is used to represent the final state. In many applictions this is not sufficient 
to provide a very good numerical accuracy. In order to allow one to include an arbitrary, in principle, number of different functions (or basis 
functions) for the final state(s) we need to modify our method. In general, applications of several basis functions to describe the final state 
and numerical computations of the cross-section with these basis functions produces a certain fear that the newly computed photoionization and 
photodetachment cross sections may exceed the actual cross section of these processes. Furthermore, it often seems that by increasing the total 
number of basis functions used one can obtain unrealistically large cross sections of photodetachment. This reason explains why the progress 
achieved in accurate numerical computations of the photodetachment cross sections is relatively modest. For brevity, in this and next Sections 
we will only discuss the process of photodetachment also although our entire analysis is also valid for the photoionization of atoms (and even 
molecules). 

By investigating this problem we have found a rigorous variational principle that always guarantees one-sided convergence of the computed 
cross sections to a certain limit (which is the real photodetachment cross-section) when the number of basis functions used to represent the 
final state of increases. This powerful variational principle makes all numerical computations of the photodetachment cross sections simple 
and transparent. Furthermore, with this principle accurate numerical computations of the photodetachment cross section become similar to 
analogous variational calculations of the total energies of bound states in few-electron atomic systems. To describe this variational 
principle explicitly, we need to introduce a few additional notations. In particular, below the notation $\Psi_{N}({\bf x}_1, {\bf x}_2, 
\ldots, {\bf x}_N)$ stands for the $N-$elecron wave function of the initial (bound) atomic state. Analogous notation $\Psi_{N-1}({\bf x}_2, 
\ldots, {\bf x}_N)$ designates the $(N-1)$-electron wave function of the final (bound) atomic fragment, or atomic fragment which arises (or 
forms) after photodetachment. The wave function of outgoing photoelectron is denoted by the notation $\phi({\bf r}_{1})$, while the 4-vector 
notation ${\bf x}_k = ({\bf r}_k, s_k)$ stands for a set of three Cartesian and one spin coordinates of the $k-$th electron. 

As is well known (see, e.g., \cite{Fock}, \cite{Eps}) the $\Psi_{N}({\bf x}_1, {\bf x}_2, \ldots, {\bf x}_N)$ and $\Psi_{N-1}({\bf x}_2, \ldots, 
{\bf x}_N)$ wave functions are the solutions of the two following variational, bound state problems 
\begin{eqnarray}
  E_{N} = min_{\{\Psi_{N}\}} \frac{\langle \Psi_N \mid H_{N} \mid \Psi_N \rangle}{\langle \Psi_{N} \mid \Psi_{N} \rangle} \; \; {\rm and} \; \; 
 E_{N-1} = min_{\{\Psi_{N-1}\}} \frac{\langle \Psi_{N-1} \mid H_{N-1} \mid \Psi_{N-1} \rangle}{\langle \Psi_{N-1} \mid \Psi_{N-1} \rangle} \; 
 \label{eigen}
\end{eqnarray} 
where $E_{N}$ and $E_{N-1}$ are the minimal (or extremal) values of these two functionals which depend upon the $\Psi_N$ and $\Psi_{N-1}$ trial 
functions. In numerous books on Quantum Mechanics (see, e.g., \cite{Fock}, \cite{Eps} and references therein) it is also shown that these minimal 
values of $E_{N}$ and $E_{N-1}$ coincide with corresponding eigenvalues of the bound state Hamiltonians of the $N-$ and $(N - 1)$-electron atoms, 
respectively. The trial functions $\Psi_{N}({\bf x}_1, {\bf x}_2, \ldots, {\bf x}_N)$ and $\Psi_{N-1}({\bf x}_2, \ldots, {\bf x}_N)$ which provide 
such minima coincide with (or they are very good approximations to) the exact solutions of the Schr\"{o}dinger equations: $H_{N} \Psi_{N} = E_{N} 
\Psi_{N}$ and $H_{N-1} \Psi_{N-1} = E_{N-1}  \Psi_{N-1}$. In these equations the notations $H_{N}$ and $H_{N-1}$ stand for the $N-$ and $(N - 
1)$-electron atomic Hamiltonains. These Hamiltonians are (in atomic units) 
\begin{eqnarray}
 H_{N} = \sum^{N}_{i=1} \Bigl[ \Bigl( - \frac12 \nabla^{2}_{i} - \frac{Q}{r_{i}} \Bigr) + \sum^{N}_{j=2(j>i)} \frac{1}{r_{ij}} \Bigr] \; \; 
 \label{Hamoilt1}
\end{eqnarray} 
and 
\begin{eqnarray}
 H_{N-1} = \sum^{N}_{i=2} \bigl[ \Bigl( - \frac12 \nabla^{2}_{i} - \frac{Q}{r_{i}} \Bigr) + \sum^{N}_{j=3(j>i)} \frac{1}{r_{ij}} \Bigr] \; \; 
 \label{Hamoilt2}
\end{eqnarray}

The both $\Psi_{N}({\bf x}_1, {\bf x}_2, \ldots, {\bf x}_N)$ and $\Psi_{N-1}({\bf x}_2, \ldots, {\bf x}_N)$ wave functions are square integrable. 
In contrast with this, the wave function of the final photoelectron $\phi({\bf x}_1)$ can be normalized only to the corresponding delta-function, 
since this one-elecron wave function does not represent any bound state. This fact is well known and interesting, but for the purposes of this 
study it is crucial to note that the two following integrals 
\begin{eqnarray}
 s = \int_{1}  \int_{2} \ldots \int_{N} \phi({\bf x}_1) \Psi_{N-1}({\bf x}_2, \ldots, {\bf x}_N) ({\bf e} \nabla_{1}) \Psi_{N}({\bf x}_1, 
 {\bf x}_2, \ldots, {\bf x}_N) d{\bf x}_1 d{\bf x}_2 \ldots  d{\bf x}_N \; \; \label{s}
\end{eqnarray} 
and 
\begin{eqnarray}
 t = \int_{1} \int_{2} \ldots \int_{N} \phi({\bf x}_1) \Psi_{N-1}({\bf x}_2, \ldots, {\bf x}_N) \Psi_{N}({\bf x}_1, {\bf x}_2, \ldots, {\bf 
 x}_N) d{\bf x}_1 d{\bf x}_2 \ldots d{\bf x}_N \; \; \label{t}
\end{eqnarray} 
do exist, and each of them is a finite number (here the notation ${\bf e}$ stands for an arbitrary unit vector in three-dimensional space).
Based on this fact, we will replace the original problem with an equivalent variational problem. To avoid a pure formal discussion let us 
consider the following simple example. Suppose we chose the wave function of the final photoelectron which is represented in the form of the 
following sum 
\begin{eqnarray}
 \tilde{\phi}({\bf x}_{1}) &=& \phi({\bf x}_{1}) \Bigl[ C_1 + \frac{C_2}{r + a} + \frac{C_3}{(r + a)^{2}} + \frac{C_4}{(r + a)^{3}} + \ldots 
 \Bigr] = \phi({\bf x}_{1}) \Bigl[ \sum^{n}_{k=1} \frac{C_k}{(r + a)^{k-1}} \Bigr] \; \; \; \label{rep1} \\ 
 &=& C_1 \phi_{1}({\bf x}_{1}) + C_2 \phi_{2}({\bf x}_{1}) + \ldots + C_n \phi_{n}({\bf x}_{1}) = \sum^{n}_{k=1} C_k \phi_{k}({\bf x}_{1}) 
 \nonumber \\
 &=& C_1 \phi({\bf x}_{1}) + C_2 \frac{\phi({\bf x}_{1})}{r + a} + \ldots + C_n \frac{\phi({\bf x}_{1})}{(r + a)^{n-1}} = \sum^{n}_{k=1} C_k 
 \frac{\phi({\bf x}_{1})}{(r + a)^{k-1}} \nonumber
\end{eqnarray} 
where the $C_1, C_2, \ldots, C_n$ coefficients are the linear parameters of these expansions, while $a$ is some positive numerical constant. 
Now, we want to show that the series (or expansion), Eq.(\ref{rep1}), is, in fact, a variational expansion and the coefficients $C_1, C_2, 
\ldots, C_n$ are the linear variational parameters of this expansion. The same statement will be true for other similar expansions. 

First, we need to consider the following integrals $s_k$ and $t_k$ which are similar to the corresponding integrals Eq.(\ref{s}) and 
Eq.(\ref{s}) 
\begin{eqnarray}
 s_k = \int_{1}  \int_{2} \ldots \int_{N} \frac{\phi({\bf x}_1)}{(r + a)^{k-1}} \Psi_{N-1}({\bf x}_2, \ldots, {\bf x}_N) ({\bf e} \nabla_{1}) 
 \Psi_{N}({\bf x}_1, {\bf x}_2, \ldots, {\bf x}_N) d{\bf x}_1 d{\bf x}_2 \ldots d{\bf x}_N \; \; \label{s-k}
\end{eqnarray} 
and 
\begin{eqnarray}
 t_k = \int_{1} \int_{2} \ldots \int_{N} \frac{\phi({\bf x}_1)}{(r + a)^{k-1}} \Psi_{N-1}({\bf x}_2, \ldots, {\bf x}_N) \Psi_{N}({\bf x}_1, 
 {\bf x}_2, \ldots, {\bf x}_N) d{\bf x}_1 d{\bf x}_2 \ldots d{\bf x}_N \; \; \label{t-k}
\end{eqnarray} 
where $k$ is the index of the radial $k-$th basis function of the final photoelectron in Eq.(\ref{rep1}). The one-electron function 
$\phi({\bf x}_1)$ coincides with the wave function of the photoelectron, defined either by Eq.(\ref{rad1}), or by Eq.(\ref{rad2}) mentioned 
above. This means that all integrals $s_k$ and $t_k$ in Eqs.(\ref{s-k}) and (\ref{t-k}) exist and they are finite.

Second, by using these integrals we can construct the following functional $F$, which is a quadratic form upon the linear coefficients $C_1, 
C_2, \ldots, C_n$, and can be called the photodetachment functional \cite{GF}. This photodetachment functional $F$ is written in a very compact 
form by using the row vector ${\bf C}^{\dagger} = (C_1, C_2, \ldots, C_n)$ and its conjugate column vector ${\bf C} = ({\bf 
C}^{\dagger})^{\dagger}$:  
\begin{eqnarray}
 F = {\bf C}^{\dagger} \Bigl( \hat{S} - \sigma \hat{T} \Bigr) {\bf C} = \sum^{n}_{i=1} \sum^{n}_{j=1} \Bigl( \hat{S}_{ij} - \sigma \hat{T}_{ij} 
 \Bigr) C_i C_j = \sum^{n}_{i=1} \sum^{n}_{j=1} \Bigl( s_i s_j - \sigma t_i t_j \Bigr) C_i C_j \; \; \label{ff}
\end{eqnarray} 
where the $n \times n$ matrices $\hat{S}$ and $\hat{T}$ are the $n-$dimensional diadas \cite{Kochin}, i.e., they are the two symmetric matirces 
which have the following matrix elements: $S_{ij} = s_{i} s_{j}$ and $T_{ij} = t_{i} t_{j}$, respectively. Here we assume that all components 
of the vector ${\bf C}$ are real. The optimal coefficients $C_1, C_2, \ldots, C_n$ form the vector which is the solution of the following 
generalized eigenvalue problem 
\begin{eqnarray}
  \Bigr( \hat{S} - \sigma \hat{T} \Bigr) {\bf C} = 0 \; \; \label{eigenEq}
\end{eqnarray}
where $\sigma$ is the eigenvalue of this equation which also is the maximum (or minimum) of the functional $F$, Eq.(\ref{ff}), or photodetachment 
functional $F$ (see, e.g., \$ 17 in \cite{GelLA}). It is also clear that the numerical value of $\sigma$ coinsides with the actual cross section of 
photodetachment, which also is the maximum (or minimum) of the photodetachment functional $F$. On the other hand, optimization of the functional 
$F$ is completely equivalent \cite{GF} to the obtaining of optimal projection of the $n$-dimensional vector ${\bf s} = (s_1, s_2, \ldots, s_n)$ on 
the $n$-dimensional vector ${\bf t} = (t_1, t_2, \ldots, t_n)$. The last problem is considered in detail in \$ 3 of \cite{GelLA} and it is usually 
solved by the method of least squares. Thus, we have shown that the wave function of the final state for atomic photodetachment is either already 
optimal, or can be made so during the optimization process. In fact, without loss of generality we can always assume that the wave function of the 
final state for atomic photodetachment process is optimal. Such a conclusion leads to a number of very interesting general facts about actual 
physical states and processes which are briefly considered below.   

\subsection{On the optimal nature of actual physical states}

Let us note that our variational mehtod described above can be generalized to other transition processes which are studied in the both non-relativistic 
Quantum Mechanics and Quantum Electrodynamics. Here we want to show this explicitly. Everywhere below, we shall always assume that the probability of a 
transition (or transition probability) during any given physical process is determined by the Fermi's golden rule, i.e., the probability of this process 
is proportional to the square of the transition amplitude. This rule is written in the following general form \cite{Fermi}
\begin{equation}
 P_{i \rightarrow f} =  2 \pi {\cal C} \mid \langle \Psi_{fi} \mid \hat{A} \mid \Psi_{in} \rangle \mid^{2} = 2 \pi {\cal C} \mid \langle 
 \Psi_{fi} \mid \hat{A} \mid \Psi_{in} \rangle \langle \Psi_{in} \mid \hat{A} \mid \Psi_{fi} \rangle \mid \; \; , \; \label{prob}
\end{equation} 
where $\hat{A}$ is the quantum operator which represent a given process, while ${\cal C}$ is some positive numerical constant which depends only upon 
the process itself and units used. All calculations of the transition amplitude are performed for the finite, spatial $3D-$volumes, $V$ which can be 
very large, but always finite. Now, by using the Fermi's golden rule and our method described above one can show that for any given wave function of 
the initial state $\Psi_{in}$, the wave function of the final state $\Psi_{fi}$ is either already optimal, or can be made so during some optimization 
process. Furthermore, it is possible to show that for any given wave function of the final state $\Psi_{fi}$, the wave function of the initial state 
$\Psi_{in}$ is either already optimal, or can be made so during another optimization process. In other words, if we are using the Fermi's golden rule 
to determine probability of the transition process we can restrict ourselves to the optimial initial and states only, i.e., to the states which have 
the optimial wave functions for this process. It is clear that the chosen optimization procedure unambigously depends upon the operator $\hat{A}$ only, 
or in other words, upon the given physical process.  

Now it seems very tempting to identify the optimal wave functions with the wave functions of actual physical systems. In this case, the optimal states 
themselves must be indentified with actual (or physical) states of these systems. Remarcably, but this works in reality and this fact allows us to say 
that every physical state in Nature is always optimal for a given physical process (as well as the result of this process). Also, without loss of 
generality, we can assume that the Fermi's golden rule always describes transitions between two optimal (or physical) states in one physical system. 
To emphasize the uniqueness and importance of the Fermi's golden rule, we recommend that the readers themself deal with the cases when the transition 
probability is determined by some other formulas, e.g., 
\begin{equation}
 P_{i \rightarrow f} =  2 \pi {\cal C} \mid \langle \Psi_{fi} \mid \hat{A} \mid \Psi_{in} \rangle \mid^{n} \; \; , \; \label{probwr}
\end{equation} 
where $n$ = 1, 3 and 5. In such processes one easily finds some non-optimal final states (or ghost states), which cannot be optimized, in principle, 
and cannot be considered as `physical' states. Cross sections determined with such states can be made as large (or small) as you want.  
 
For our present purposes it is important to note that for any given transition process from a certain initial state the Nature always choses the 
optimal (and unique) final state. This final state provides the maximal (or minimal) projection (for this process) on the given initial state. 
Therefore, all cross sections, which are determined by the Fermi's golden rule, have extremal properties. In other words, these cross sections are 
always maximal (or minimal) values of some functional (in our case this is the photodetachment functional $F$) in respect to all possible variations 
of the trial functions of the final states. In principle, future variational computations of the cross sections of different transition processes may 
fill atomic quantum mechanics and quantum electrodynamics with new variational content and make the process of calculating of such cross sections more 
competative, complete and accurate.

\section{Discussion and Conclusions}  

We have studied the non-relativistic photodetachment of the outer-most electrons in different atomic systems including neutral atoms and positively 
charged atomic ions which contain $N_e$ (where $N_e \ge 2$) bound electrons in an atom/ion with the nuclear charge $Q$. Photodetachment of the 
outer-most electron in one-electron atoms/ions and negatively charged atomic ions, where $N_e = Q + 1$, is also investigated. In each of these cases 
we have derived the closed analytical formulas for the both differential and total cross sections of photoionization and/or photodetachment (see 
Eqs.(\ref{dsigmaZpol}), (\ref{sigmaZ}), (\ref{sigma-ang}) and (\ref{sigma-totl}) above). These formulas are the main results of this study and each 
of them contains only a few basic parameters of the problem, e.g., the cyclic frequency of incident light $\omega$, atomic ionization potential $I$, 
the total number of initially bound electrons $N_e$ and electrical charge of atomic nucleus $Q$. For neutral atoms and positively charged ions with 
$N_e \ge 2$ and for negaively charged ions similar formulas have never been produced in earlier papers. Our procedure developed to determine the 
differential and total cross sections of photoionization of arbitrary few- and many-electron atomic systems. This can be considered as a substantial 
improvement of earlier method proposed in \cite{BurSeat} for few- and many-electron atoms and ions. Note that this method from \cite{Bates} and 
\cite{BurSeat} is still extensively used to explain many qualitative features known for photoionization of various few- and many-elecron atoms and 
develop different approximate formulas for the cross sections. 
   
Our current method was also tested in applications to photoionization of the ground state(s) in one-electron atomic systems. Analytical formula for 
the differential and total cross sections derived for one-electron atomic systems coincides with the well known formula obtained in \cite{Stob}. 
Derivation of analogous formulas to photoionization of the excited state(s) in one-electron atomic systems is discussed in the Appendix. Our 
analytical formulas for the photodetachment cross sections of the negatively charged ions are original and complete. Moreover, these our formulas 
include only elementary functions. None of these formulas has ever been produced in earlier studies. In general, the cross sections of 
photodetachment of all negatively charged ions can be determined from numerical computations. However, based on columns of numbers with many digits 
in each it is very hard (even impossible) to predict analytical formulas which lead to these results. By using this (numerical) approach only you 
will always miss something important and interesting.  

By using our formulas derived above we have determined the cross-sections of photoionization and photodetachment of some systems discussed in this 
study. Numerical results can be found in Table I. Table I contains photodetachment cross sections of the negatively charged H$^{-}$ ion (its ground 
singlet $1^{1}S(L = 0)-$state) and neutral hydrogen atom H (the ground doublet $1^{2}s(\ell = 0)$-state). In this Table the cross section are 
considered as a function of cyclic frequency $\omega$ of the incident photons (here $\omega \ge I$. These frequencies are shown in atomic units of 
energy $\hbar \omega$. To express these values in $GigaHetrz$ (of $GHz$, for short) one can use the relation 1 $a.u.$ = 6.57968392050160$\cdot 
10^{6}$ GHz. This conversion factor is recomended by NIST (2020) to be used in scientific research). The observed agreement between our data and 
results known earlier studies is excelent for the neutral hydrogen atom and very good for the negatively charged H$^{-}$ ion, but for large and 
very large frequencies actual agreement is not very good.        

In general, our analytical formulas for the photoionization and photodetachment cross sections and other results derived in this study are of 
interest in many problems which currently exist in stellar astrophysics, physics of high- and low-temperature plasmas of light elements, solid 
state physics and other areas of natural science. In the course of our investigation we have also discovered an important principle of the optimial 
final state which allows one to conduct accurate and even highly accurate computations of photoionization and photodetachment cross-sections by 
using a number of different basis functions which approximate (or represent) the final state wave function.    

Note that in this study our analysis was restricted to photodetachment of the outer-most electrons only. Applications of our method are mainly 
related to calculations of actual opacities of the Solar and stellar photospheres. In respect to this restriction, we did not consider 
photoionization of electrons from internal electron shells, double photoionization, photodetachment with instantaneous electron(s) bound-bound 
transitions and other processes. In other words, in this study we deal with the low-energy photoionization and/or photodetachment only. Relation 
between our current low-energy analysis and actual experiments can be understood from the literature (see, e.g., \cite{Fro2015}, \cite{Star}, 
\cite{Kry}, \cite{Peg} and references therein). Photodetachment of the outer-most electrons in light atomic systems is very common in nature. On 
the other hand, the corresponding cross-sections can be determined analytically and presented in the form of closed formulas. This allows one to 
create a reliable basis for accurate numerical calculations of the photoionization and photodetachment cross-sections for more complicated systems, 
since numerical results can be compared with the known theoretical predictions. \\

{\bf Appendix} 

Let us consider photoionization of one-electron atoms/ions from their excited states. In these cases the expression, Eq.(\ref{rad3}), for the 
electron density of the incident ground state cannot be used. Instead, we have to deal with the electron (radial) wave function written in the 
form 
\begin{eqnarray}
 R_{n \ell}(r) = - \frac{2 Q}{n^2} \sqrt{\frac{Q (n - \ell - 1)!}{[(n + \ell)!]^3}} \Bigl(\frac{2 Q r}{n}\Bigr)^{\ell} 
 L^{2\ell+1}_{n+\ell}\Bigl(\frac{2 Q r}{n}\Bigr) \exp\Bigl(- \frac{Q r}{n}\Bigr) \; \; \label{radfun1eA}
\end{eqnarray}
where $Q$ is the electric charge of the nucleus, while $n$ and $\ell$ are the corresponding quantum numbers. The notation $L^{m}_{n}(x)$ stands 
for the generalized Laguerre polynomials, i.e., 
\begin{eqnarray}
  L^{m}_{n}(x) &=& (-1)^{m} \frac{n!}{(n - m)!} \exp( x ) x^{-m} \frac{d^{n-m}}{d x^{n-m}} \Bigl[ \exp(- x) x^{n} \Bigr] = 
 \sum^{n-m}_{k=0} (-1)^{k} \frac{n!}{(n - m - k)!} \frac{x^{k}}{k!} \; \; \nonumber \\
 &=& \frac{n!}{(n - m)!} \sum^{n-m}_{k=0} (-1)^{k} C^{k}_{n-m} \frac{x^k}{k!} = m! C^{m}_{n} \sum^{n-m}_{k=0} (-1)^{k} C^{k}_{n-m} 
 \frac{x^k}{k!} \; \; \label{Laug} 
\end{eqnarray}
where the notation $C^{p}_{q}$ designates the binomial coefficients, or number of combinations from $q$ by $p$. Here the both $p$ and $q$ are 
integer numbers and $p \le q$.  

The radial derivative of the wave function $R_{n \ell}(r)$ of the initial (bound) state is  
\begin{eqnarray}
 \frac{\partial}{\partial r} R_{n \ell}(r) &=& \frac{(2 Q)^2}{n^3} \sqrt{\frac{Q (n - \ell - 1)!}{[(n + \ell)!]^3}} \Bigl[ \Bigl( 1 -  
 \frac{n}{2 Q r} \Bigr) \Bigl(\frac{2 Q r}{n}\Bigr)^{\ell} L^{2\ell+1}_{n+\ell}\Bigl(\frac{2 Q r}{n}\Bigr) 
 \nonumber \\
 &+& \Bigl(\frac{2 Q r}{n}\Bigr)^{\ell} L^{2\ell+2}_{n+\ell-1}\Bigl(\frac{2 Q r}{n}\Bigr) \Bigr] \exp\Bigl(- \frac{Q r}{n}\Bigr)
 \; \; \label{radfun2eA}
\end{eqnarray} 
This formula must be multiplied by the wave function of the final, unbound electron (multiplied by a factor $\frac{2 \ell_1 + 1}{2 p}\Bigr)$ 
which must be taken in the form 
\begin{eqnarray}
 &&R_{\ell_1;p}(r) = \Bigl(\frac{2 \ell_1 + 1}{2 p}\Bigr) \frac{2^{\ell_1} Q p^{\ell_1}}{(2 \ell_1 + 1)!} \sqrt{\frac{8 \pi (1 + 
 \nu^{2})}{\nu [1 - \exp(-2 \pi \nu)]}} \; \; r^{\ell_1} \exp(-\imath p r) \; \; {}_1F_{1}(\ell_1 + 1 + \imath \nu,  2 \ell_1 + 2; \nonumber \\
 &&  2 \ell_1 + 2; 2 \imath p r) = \frac{2^{\ell_1} p^{\ell_1}}{(2 \ell_1)!} \sqrt{\frac{2 \pi \nu (1 + \nu^{2})}{1 - \exp(-2 \pi  \nu)}} \; \; 
 r^{\ell_1} \exp(-\imath p r) \; \; {}_1F_{1}(\ell_1 + 1 + \imath \nu, 2 \ell_1 + 2; 2 \imath p r) \; \; \nonumber
\end{eqnarray}
where the notation $\nu$ stands for the parameter $\nu = \frac{Q}{p}$, which was used in Sections III and IV above, while the momentum of 
outgoing photoelectron $\ell_1$ can be equal either $\ell + 1$, or $\ell - 1$ (if $\ell \ge 1$). This selection rule applies to 
photoionization of one-electron atoms and/or ions in the non-relativistic dipole approximation. The radial function $R_{\ell_1;p}(r)$ 
eesentially coinsides with the regular Coulomb wave function $F(\eta, p r)$ which is multiplied by the factor $\frac{1}{p r}$ (in this 
study we also multiply this function by an additional factor $\frac{2 \ell_1 + 1}{2 p})$.  

At the last step of our method we have to determine the following radial integral 
\begin{eqnarray}
  I_{rd} =  I_{rd}(n, \ell, p, \ell_1) = \int_{0}^{+\infty} R_{\ell_1;p}(r) \Bigl[ \frac{\partial}{\partial r} R_{n \ell}(r) \Bigr] 
  r^2 dr \; \; . \; \label{RadInt}
\end{eqnarray}
To derive the explicit expression for this radial integral and obtain analytical formulas for the photoionization cross sections one needs 
to use the following general formula (see, e.g., \cite{GR})  
\begin{eqnarray}
 \int^{+\infty}_{0} {}_{1}F_{1}(a, c; kt) t^{b-1} \exp(-s t) dt = \frac{\Gamma(b)}{s^{b}} \; {}_{2}F_{1}(a, b; c; \frac{k}{s} \Bigr) 
 = \frac{\Gamma(b)}{(s - k)^{b}} \; {}_{2}F_{1}(c - a, b; c; \frac{k}{k - s} \Bigr) \; \; \label{IntHGF}
\end{eqnarray}
Now, it is easy to see that our original problem is reduced to analytical and/or numerical calculations of the Lagurre polynomials of 
the (2,1)-hypergeometric functions. In other words, each of the the radial intergal, Eq.(\ref{RadInt}), a Lagurre polynomial which 
contain a number of different (2,1)-hypergeometric functions (as their arguments). In general, such analytical computations are 
relatively simple, but the final formula is cumbersome.

\begin{table}[tbp]
   \caption{Photodetachment cross sections (in $cm^{2}$) of the negatively charged hydrogen ion H$^{-}$ and neutral H atom 
            for different frequencies $\omega$ (in atomic units).} 
     \begin{center}
     \scalebox{0.65}{%
     \begin{tabular}{| c | c | c | c | c | c | c | c |}
           \hline\hline
 $\omega$  & $\sigma({\rm H}^{-})$ & $\omega$ & $\sigma({\rm H}^{-})$ & $\omega$ & $\sigma({\rm H})$ & $\omega$ & $\sigma({\rm H})$ \\  
           \hline\hline 
        0.02900   &   0.177823402579E-17   &    0.07900&  0.402048343777E-16 &    0.50125   &   0.187874280218E-16   &    0.55125&  0.145611577140E-16 \\
        0.03025   &   0.449903332178E-17   &    0.08025&  0.402429509797E-16 &    0.50250   &   0.186630134378E-16   &    0.55250&  0.144726043460E-16 \\
        0.03150   &   0.743164374224E-17   &    0.08150&  0.402723784180E-16 &    0.50375   &   0.185396982200E-16   &    0.55375&  0.143847679366E-16 \\
        0.03275   &   0.103367259770E-16   &    0.08275&  0.402938193317E-16 &    0.50500   &   0.184174701603E-16   &    0.55500&  0.142976412018E-16 \\
        0.03400   &   0.131080384615E-16   &    0.08400&  0.403079172491E-16 &    0.50625   &   0.182963172139E-16   &    0.55625&  0.142112169466E-16 \\ 
        0.03525   &   0.156983882506E-16   &    0.08525&  0.403152620454E-16 &    0.50750   &   0.181762274973E-16   &    0.55750&  0.141254880642E-16 \\
        0.03650   &   0.180903895426E-16   &    0.08650&  0.403163948487E-16 &    0.50875   &   0.180571892856E-16   &    0.55875&  0.140404475345E-16 \\
        0.03775   &   0.202824419475E-16   &    0.08775&  0.403118124543E-16 &    0.51000   &   0.179391910097E-16   &    0.56000&  0.139560884230E-16 \\
        0.03900   &   0.222813335219E-16   &    0.08900&  0.403019713021E-16 &    0.51125   &   0.178222212544E-16   &    0.56125&  0.138724038793E-16 \\
        0.04025   &   0.240981305064E-16   &    0.09025&  0.402872910616E-16 &    0.51250   &   0.177062687556E-16   &    0.56250&  0.137893871363E-16 \\
        0.04150   &   0.257458310108E-16   &    1.68650&  0.188721335725E-16 &    0.51375   &   0.175913223981E-16   &    2.15875&  0.293589939513E-18 \\
        0.04275   &   0.272380141786E-16   &    1.75150&  0.186240294038E-16 &    0.51500   &   0.174773712135E-16   &    2.22375&  0.268553401977E-18 \\
        0.04400   &   0.285880698101E-16   &    1.81775&  0.183823024187E-16 &    0.51625   &   0.173644043773E-16   &    2.29000&  0.245842609892E-18 \\
        0.04525   &   0.298087734171E-16   &    1.88525&  0.181467092367E-16 &    0.51750   &   0.172524112075E-16   &    2.35750&  0.225227248542E-18 \\
        0.04650   &   0.309120690042E-16   &    1.95400&  0.179170189678E-16 &    0.51875   &   0.171413811617E-16   &    2.42625&  0.206500795053E-18 \\
        0.04775   &   0.319089768968E-16   &    2.02400&  0.176930123865E-16 &    0.52000   &   0.170313038354E-16   &    2.49625&  0.189478020231E-18 \\
        0.04900   &   0.328095761149E-16   &    2.09525&  0.174744811746E-16 &    0.52125   &   0.169221689598E-16   &    2.56750&  0.173992746826E-18 \\
        0.05025   &   0.336230301061E-16   &    2.16775&  0.172612272256E-16 &    0.52250   &   0.168139663996E-16   &    2.64000&  0.159895840469E-18 \\
        0.05150   &   0.343576364754E-16   &    2.24150&  0.170530620047E-16 &    0.52375   &   0.167066861509E-16   &    2.71375&  0.147053411045E-18 \\
        0.05275   &   0.350208887044E-16   &    2.31650&  0.168498059583E-16 &    0.52500   &   0.166003183396E-16   &    2.78875&  0.135345203827E-18 \\
        0.05400   &   0.356195424683E-16   &    2.39275&  0.166512879698E-16 &    0.52625   &   0.164948532188E-16   &    2.86500&  0.124663161347E-18 \\
        0.05525   &   0.361596820876E-16   &    2.47025&  0.164573448560E-16 &    0.52750   &   0.163902811677E-16   &    2.94250&  0.114910138540E-18 \\
        0.05650   &   0.366467845037E-16   &    2.54900&  0.162678209003E-16 &    0.52875   &   0.162865926888E-16   &    3.02125&  0.105998755304E-18 \\
        0.05775   &   0.370857793488E-16   &    2.62900&  0.160825674208E-16 &    0.53000   &   0.161837784067E-16   &    3.10125&  0.978503720370E-19 \\
        0.05900   &   0.374811044209E-16   &    2.71025&  0.159014423688E-16 &    0.53125   &   0.160818290660E-16   &    3.18250&  0.903941751375E-19 \\
        0.06025   &   0.378367563353E-16   &    2.79275&  0.157243099554E-16 &    0.53250   &   0.159807355295E-16   &    3.26500&  0.835663607178E-19 \\
        0.06150   &   0.381563364071E-16   &    2.87650&  0.155510403046E-16 &    0.53375   &   0.158804887764E-16   &    3.34875&  0.773094059812E-19 \\
        0.06275   &   0.384430919782E-16   &    2.96150&  0.153815091290E-16 &    0.53500   &   0.157810799008E-16   &    3.43375&  0.715714187889E-19 \\
        0.06400   &   0.386999534954E-16   &    3.04775&  0.152155974283E-16 &    0.53625   &   0.156825001098E-16   &    3.52000&  0.663055569430E-19 \\
        0.06525   &   0.389295676823E-16   &    3.13525&  0.150531912072E-16 &    0.53750   &   0.155847407215E-16   &    3.60750&  0.614695096010E-19 \\ 
        0.06650   &   0.391343271600E-16   &    3.22400&  0.148941812117E-16 &    0.53875   &   0.154877931642E-16   &    3.69625&  0.570250340573E-19 \\
        0.06775   &   0.393163968652E-16   &    3.31400&  0.147384626824E-16 &    0.54000   &   0.153916489738E-16   &    3.78625&  0.529375418496E-19 \\
        0.06900   &   0.394777375934E-16   &    3.40525&  0.145859351233E-16 &    0.54125   &   0.152962997930E-16   &    3.87750&  0.491757288110E-19 \\
        0.07025   &   0.396201269736E-16   &    3.49775&  0.144365020855E-16 &    0.54250   &   0.152017373692E-16   &    3.97000&  0.457112442736E-19 \\
        0.07150   &   0.397451781544E-16   &    3.59150&  0.142900709636E-16 &    0.54375   &   0.151079535532E-16   &    4.06375&  0.425183951555E-19 \\ 
        0.07275   &   0.398543564540E-16   &    3.68650&  0.141465528049E-16 &    0.54500   &   0.150149402977E-16   &    4.15875&  0.395738811348E-19 \\ 
        0.07400   &   0.399489942020E-16   &    3.78275&  0.140058621298E-16 &    0.54625   &   0.149226896557E-16   &    4.25500&  0.368565575325E-19 \\
        0.07525   &   0.400303039769E-16   &    3.88025&  0.138679167633E-16 &    0.54750   &   0.148311937791E-16   &    4.35250&  0.343472228986E-19 \\
        0.07650   &   0.400993904198E-16   &    3.97900&  0.137326376758E-16 &    0.54875   &   0.147404449172E-16   &    4.45125&  0.320284286307E-19 \\
        0.07775   &   0.401572607862E-16   &    4.07900&  0.135999488338E-16 &    0.55000   &   0.146504354156E-16   &    4.55125&  0.298843082477E-19 \\
       \hline\hline 
  \end{tabular}}
  \end{center}
  \end{table}

\begin{thebibliography}{99} 

\bibitem{Star} A.F. Starace, \textit{Theory of Atomic Photoionization} (a Chapter contributed in: \textit{Springer Handbook of Atomic, Molecular, 
and Optical Physics} (Springer Verlag, New York - Berlin (2006))). 

\bibitem{Sobelman} I.I. Sobelman, \textit{Introduction to the Theory of Atomic Spectra} (Science, Moscow, 1977) [in Russian]. 

\bibitem{BS} H.A. Bethe and E.E. Salpeter, \textit{Quantum Mechanics of One- and Two-Electron Atoms} (Dover Publ. Inc., Mineola (NY) (2008)). 

\bibitem{Zat} O. Zatsarinny and S.S. Tayal, Phys. Rev. A {\bf 81}, 043423 (2010).

\bibitem{Peg} D.J. Pegg, Nuclear Instruments and Methods in Physics Research B {\bf 99}, 140 (1995). 

\bibitem{Kry} C.M. Oanaa and A.I. Krylov, J. Chem. Phys. {\bf 131}, 124114 (2009).

\bibitem{Ajm} M.P. Ajmera and K.T. Chung, Phys. Rev. A {\bf 12}, 475 (1975).

\bibitem{FroF} C. Froese Fischer, T. Brage and P. J\"{o}nsson, \textit{Computational Atomic Structure. 
               An MCHF Approach} (IOP Publishing, Bristol (1997)), Chpt. 10.

\bibitem{Frit} S. Fritzsche, Comp. Phys. Commun. {\bf 240}, 1 (2019).

\bibitem{Stob} B.M. Stobbe, Ann. der Phys. {\bf 7}, 682 (1930). 

\bibitem{Rodes} R. Rhodes, \textit{Dark Sun. The Making the Hydrogen Bomb} (Simon \& Shuster Paperback, New York (2005)). 

\bibitem{Bates} D.R. Bates and A. Damgaard, Phil. Trans. {\bf 242}, 101 (1949). 

\bibitem{BurSeat} A. Burgess and M. Seaton, Rev. Mod. Phys. {\bf 30}, 992 (1958).  

\bibitem{Osten} M. Hoffmann-Ostenhoff and T. Hoffmann-Ostenhoff, Phys. Rev. A \textbf{16}, 1782 (1977). 

\bibitem{Marl} R. Poeckert and J.M. Marlborough, Astrophysical Journal {\bf 218}, 220 (1977).  

\bibitem{Fro2015A} A.M. Frolov, European Physical Journal D {\bf 69}, 132 (2015).  

\bibitem{Sob} V.V. Sobolev, \textit{The Course of Theoretical Astrophysics} (Nauka, Moscow, (1967) (in Russian)). 

\bibitem{Aller} L.H. Aller, \textit{The Atmospheres of the Sun and Stars} (2nd edn., Ronald Press, New York, (1963)). 

\bibitem{Motz} L. Motz, \textit{Astrophysics and Stellar Structure} (Ginn and Company, Waltham, MA (1970)). 

\bibitem{Zirin} H. Zirin, \textit{The Solar Atmospheres} (Ginn and Company, Waltham MA (1966)).

\bibitem{AB} A. Akhiezer and V.B. Berestetskii, \textit{Quantum Electrodynamics} (4th ed., Science, Moscow, 1981) [in Russian]. 

\bibitem{IzZub} C. Itzykson and J.-B. Zuber, \textit{Quantum Field Theory} (McGraw-Hill, New York, 1980).

\bibitem{Grein} W. Greiner and J. Reinhardt, \textit{Quantum Electrodynamics} (4th ed., Springer Verlag, Berlin, 2009). 

\bibitem{BD1} A.K. Bhatia, R.J. Drachman, Phys. Rev. A {\bf 32}, 3745 (1985). 

\bibitem{GR} I.S. Gradstein and I.M. Ryzhik, \textit{Tables of Integrals, Series and Products} (6th revised ed., Academic Press, New York (2000)).

\bibitem{AS} \textit{Handbook of Mathematical Functions} (M. Abramowitz and I.A. Stegun (Eds.), Dover, New York, (1972)). 

\bibitem{Fock} V.A. Fock, \textit{Foundations of Quantum Mechanics} (2nd. ed., Nauka (Science), Moscow, 1976). 

\bibitem{LLQ} L.D. Landau and E.M. Lifshitz, \textit{Quantum Mechanics. Non-relativistic Theory} (4th ed., Pergamon Press, Oxford, 1975). 

\bibitem{BLP} V.B. Berestetskii, E.M. Lifshitz and L.P. Pitaevskii, \textit{Relativistic Quantum Theory} (Pergamon Press, Oxford, (1971)). 

\bibitem{Fro2015} A.M. Frolov, Chem. Phys. Lett., {\bf 626}, 49 (2015). 

\bibitem{Bet2} H. Bethe and C. Longmire, Phys. Rev. {\bf 77}, 647 (1950).

\bibitem{Kochin} N.E. Kochin, \textit{Vector Calculus and the Principles of Tensor Calculus} (USSR Acad. of Sciences Publishing, 9-th ed., 
Moscow, (1965)), [in Russian] Chpt. III.

\bibitem{Jacks} J.D. Jackson, \textit{Classical Electrodynamics}, (J. Wiley \& Sons Inc., New York (1975)).

\bibitem{Edm} A.R. Edmonds, \textit{Angular Momentum in Quantum Mechanics} (Princeton University Press, Princeton (1974)). 

\bibitem{Chan1} S. Chandrasekhar, Astrophys. Jour. {\bf 102}, 223 (1945).  

\bibitem{Chan2} S. Chandrasekhar, \textit{Selected Papers. Radiative Transfer and Negative Ion of Hydrogen} (University of Chicago Press, Chicago, 
1989), Vol. 2.  

\bibitem{John} T.L. John, Monthly Notices {\bf 121}, 41 (1960).

\bibitem{BetPei} H. Bethe and R. Peierls, Proc. Roy. Soc. {\bf 148}, 146 (1935). 

\bibitem{Eps} S.T. Epstein, {\it The Variation Method in Quantum Chemistry} (Academic Press, New York (1974)). 

\bibitem{GF} I.M. Gelfand and S.V. Fomin, \textit{Calculus of Variations} (Dover Publ. Inc., Mineola, New York, (1990)). 

\bibitem{GelLA} I.M. Gelfand, \textit{Lectures on Linear Algebra} (Dover Publ. Inc., New York, (1989)).

\bibitem{Fermi} E. Fermi, \textit{Nuclear Physics} (University of Chicago Press, Chcago, IL (1950)). 

\end{thebibliography}
\end{document}